# Contact and first layer residues prediction in protein dimers using the Gaussian Network model with adjustable number of fast modes

Ognjen Perišić[1]

(December 28, 2013)

[1] Current Address:
  Vojvode Brane 32,
  11000 Belgrade, Serbia
  Email address: ognjen.perisic@gmail.com

## Abstract

Interactions between proteins are hard to decipher. Protein-protein interactions are difficult problem to address because they are not based on differences in charge type like protein-DNA or protein-lipid interactions. In this manuscript we present a few methods aimed at recognizing contact and their neighboring first layer residues. The methods are based on the simplified analysis of normal modes via the Gaussian Network Model (GNM). The methods adjust the number of modes used in analysis on the basis of the comparison of predicted and expected number of targets (contact and first layer residues). The number of modes used in weighted sum calculation is increased or decreased if the number of predictions falls out of range for a given protein sequence length. The methods are able to recognize more than 50% of targets on average. At the same time they generate slightly more than 40% of false positives. They are also fairly successful in recognizing near native decoys of the Vakser decoy sets. The results depicted here show that kinetically active residues are important in protein-protein interactions. The GNM analysis depicted also shows that the two interacting chains usually exhibit opposite behavior. While the binding surface of one of the partners is rigid and stable, the interfacial area of the other partner is more flexible.

## Introduction

Protein-protein interactions are still not fully understood. They are more difficult to decipher than protein-DNA or protein-RNA interactions because they are, usually, not based on the attractive electrostatic interactions between highly negatively charged nucleic acid chains (phosphate ions) and positively charged patches on the surface of proteins. Many attempt were made to recognize the binding mechanism between individual protein chains. Here we present out attempt at recognizing protein-protein interactions. It is based on the Gaussian Network Model developed by Demirel at al [1]

The paper starts with a short overview of the theory of normal mode analysis using the Gaussian network model (GNM) [1]. After that, we describe what we are trying to decipher, i.e. what are our targets. The description of the GNM code is given after that. The first simple 1D prediction (sequential neighbors influence) based on five fastest modes only is given in the third chapter. In the same chapter, we also introduce a prediction based on fastest 10% modes. After that, we describe the behavior of dimers with different lengths of protein chains and introduce the improvement of the simple model – prediction based on the adjustable number of modes. In the next chapter, we introduce the predication improvement, i.e., the movement from the one dimensional (1D) prediction to a full 3D approach. There we first describe a simple model with fixed number of residues being influenced, and after that the variable number of residues being influences. In the next chapter we combine adjustable 1D and 3D approaches. That method offers an improvement over 3D prediction. We next test our prediction algorithms on the Vakser decoy sets. We compare our coarse-grained methods to Lu and Skolnick's detailed, residue level statistical potential approach. The paper ends with the Conclusion.

## 1. Theory

Normal modes are calculated via the $C_\alpha$ - $C_\alpha$ contact map generated for every protein chain. The modes are eigenvectors of that map (square matrix). Mode intensities (i.e. temperatures) are eigenvalues of that map. The weighted sum of the fastest modes is used to decipher their role in dimer interactions.

The Kirchhoff matrix $\Gamma$ of contacts

$$\Gamma = \begin{cases} H(r_c - r_{ij}) & i \neq j \\ -\sum_{i(\neq j)}^{N} \Gamma_{ij} & i = j \end{cases} \quad (1)$$

The cross correlation associated with the $k^{th}$ mode

$$\langle \Delta \mathbf{R}_i \cdot \Delta \mathbf{R}_j \rangle = (3k_B T / \gamma)[\lambda_k^{-1} \mathbf{u}_k \mathbf{u}_k^t]_{ij} = (3k_B T / \gamma)\lambda_k^{-1}[\mathbf{u}_k]_i [\mathbf{u}_k]_j \quad (2)$$

The weighted sum of the modes from $k_1$ to $k_2$

$$\langle (\Delta \mathbf{R}_i)^2 \rangle_{k_1 - k_2} = (k_B T / \gamma)\sum_{k_1}^{k_2} \lambda_k^{-1}[\mathbf{u}_k]_i^2 \Bigg/ \sum_{k_1}^{k_2} \lambda_k^{-1} \quad (3)$$

## 2. Targets, tools and training sets

### a. Targets

Our aim is to accurately predict the contact residues in protein dimers, and, thus, possibly, connect the structural patterns in those proteins to their contact patches. We define a contact residue as a residue in which at least one atom is at the maximum distance of 4.5 Å from one or more atoms from one or more residues on the surface of the other chain. The distance of 4.5 Å corresponds to the size of one water molecule. We also use the notion of a first layer residue (FLR). That is a neighboring residue to the chain's contact residues, belonging to that same chain (at the maximum distance of 4.5 Å). We developed a C++ code to extract the contact and first layer residues for every chain we analyzed. The program's output is the number of atoms per each contact or first layer residue, followed by its total number of atoms.

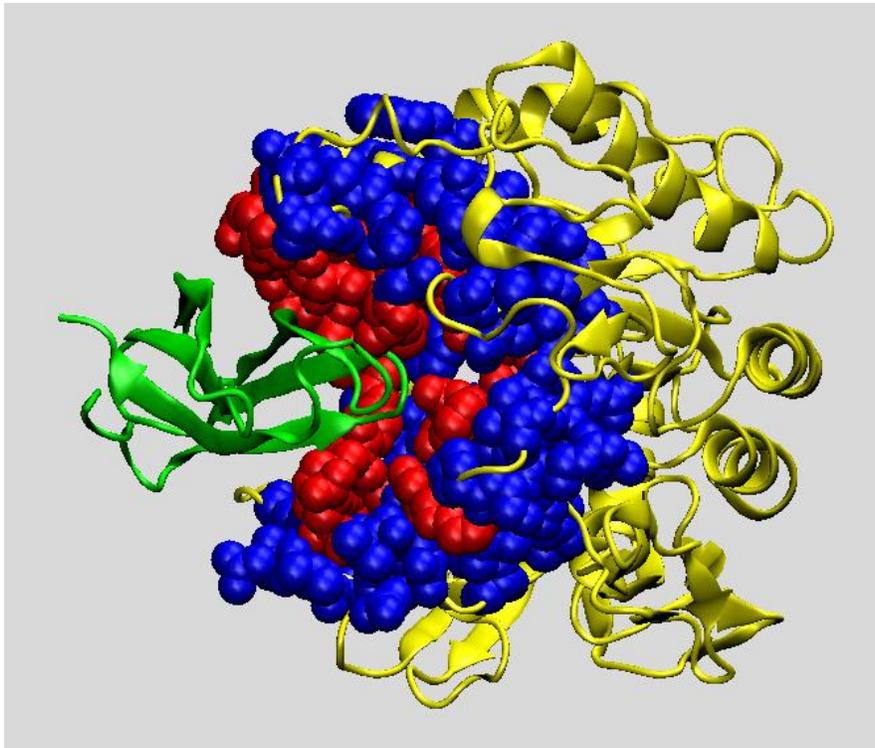

Figure 1. Visualization of the targets we aim at using the Hydrolase/hydrolase inhibitor from pig (pdb code is 1BVN). The pancreatic alpha-amylase is shown in complex with the proteinaceous inhibitor tendamistat. There are two chains, P and T. Chain P is yellow and chain T is green (both depicted cartoon representation). The chain P contact residues are colored red and visualized using VDW method. Its first layer residues (residues in direct contact with the contact residues) are colored blue and also visualized using VDW method.

### b. GNM code

The Gaussian Network Model code was also written in C++. The code first calculates the Kirchhoff contact matrix $\Gamma$ (Eq. 1). The matrix $\Gamma$ calculation is based on the distances between $C_\alpha$ atoms only, and those distances have to be less or equal to the size of one water molecule ($r_c = 4.5$ Å). The program then calculates the eigenvalues and eigenvectors for that matrix $\Gamma$. Those eigenvalues and eigenvectors are used in the weighted sum calculation (Eq. 3). Numerical procedures for eigenvalue and eigenvector calculation are taken from the book "Numerical recipes in C++"[**CITE**]. Testing and calibration were performed on the proteins used in Demirel et all [1].

### c. Training set

The training set is comprised of 414 protein dimer complexes. We separated this set into heterodimer and homodimers subsets, using two criteria: a) if the ratio of chain lengths (chain length is the number of residues in that chain) in a dimer complex is greater than 2, that complex is considered to be heterodimer; b) if the ratio of chain lengths is smaller than 2, the Smith-Waterman sequence alignment algorithm was applied to recognize and separate dimers in which chain sequences are highly similar. This approach was applied following the logic of homology modeling principles which says that high sequence similarity implies high structural similarity [**CITE**]. Therefore, the first group contains dimers in which constituents do not bear structural similarity, and the second in which members are sequentially and structurally highly similar. We did this because we assumed that behavior of heterodimers during, and upon, binding differs from the behavior of homodimers. Different behaviors of these two groups may imply that their kinetically active residues may not be have the same role in protein binding. We used this approach because the Gaussian Network Model is based on structural organization of residues, i.e., on the spatial distribution of proteins' $C_\alpha$ atoms. Of these 414 dimers, 136 are heterodimers, and the rest are homodimers. Majority of chains in our set are shorter than 300 residues, but we also have a number of chains longer than 400 residues. The distribution of chain lengths is shown in Figure A1 in the Appendix.

### 1. Simplest 1D prediction (sequential neighbors influence only) based on 5 fastest modes

The first method we tried is based on the approach of Demirel et all [1]. They used five fastest modes to recognize kinetically hot residues in proteins. We used this approach to separate the potential contact and first layer residues from the rest of residues in the chain. The first step was the calculation of the weighted sum (Eq. 3). That sum gives a kinetic contribution of each residue for that given set of modes. We normalized the sum and analyzed only hot residues with the normalized amplitude higher than 0.05. The number of hot residues is usually smaller than the number of contact or first layer residues. To account for that, we spread the influence of a hot residue to its sequential neighbors. For chains longer than 100 amino acids, we labeled the very hot residue and 4 residues upstream and four downstream from it as predictions. That means that we assumed that those residues were in contact with other chain or belong to first layer residues. We used this approach on all 414 dimers regardless the chain length, or the nature (hetero or homodimer) of a particular complex. Figure 2a depicts the example of our initial approach on four different chains. It shows the weighted sums of the four chains, their contact and first later residues (expressed as the ratio of atoms per the total number of atoms in residue) and our predictions. It is clearly visible that for the longer chains (1BVN chain P in particular), five fastest modes fail in predicting the target residues. For shorter chains (2SNI chain E, 1UDI chain E and 1CXZ chain A), five modes are better in connecting the kinetically hot residues to contact patches and FLR patches, but the overall prediction is still not very favorable because the percent of the truly predicted contact and FLR residues is comparatively small.

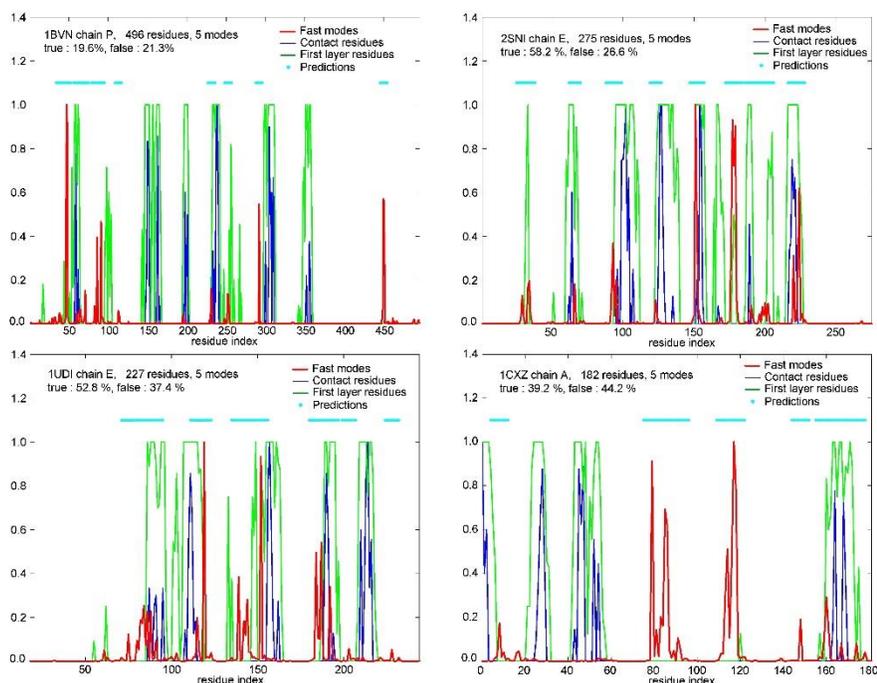

Figure 2a. An example of the 1D prediction approach in which the influence of a kinetically hot reissue (based on the 5 fastest modes only), is spread to its sequential neighbors, for 4 different chains (1BVN chain P, 2SNI chain E, 1UDI chain E and 1CXZ chain A). Red lines depict the weighted sums. Blue lines are contacts residues. Green lines are first layer residues. Cyan dots are predictions.

Figure 2b shows the algorithm output for all 828 protein chains (414 dimers). The ratio (percentage) of true predictions versus ratio of false prediction is depicted on a two dimensional Cartesian plane. The ratio of true predictions is the number of true predictions (contact and FLR) over the total number of targets. We can call it the ratio of true positives. In literature, they are also called *sensitivity*. The ratio of false predictions is the number of residues falsely predicted as being either contact or FLR over the total number of non-target residues. We call them the false positives ratio, but they can also be called *specificity*. The Cartesian plane is separated into two parts by a diagonal going from the lover left to the upper right corner. The chains that are above the diagonal are considered satisfying because the ratio of their true positives over false positives is over 1. The chains positioned under the diagonal are, obviously, unsatisfying, i.e. bad predictions. The predictions (i.e. chains) in the upper left corner are defined as good predictions. The ratio of true positives is above 0.5, and the ratio of false positives lower or equal to 0.5. In addition, we defined as very bad predictions, the ones that fall into the lower left corner (the ratio of false positives is over 0.5, and the ratio of true positives lower or equal than 0.5). From now on, we will use to this two measures, percentage of good predictions and percentage of bad predictions, besides true positives mean, and false positives mean, as measures of the quality of our prediction methods. We could use some other definitions of good and bad predictions, but we accepted the ones described above primarily for the algorithm tuning purposes. We could put, for example, chains with the ratio of true positives vs. false positives higher than 2 as good predictions, but that would put chains with 10 % of true positives and 5 % of false positives into the group of good predictions, which is, obviously, wrong.

It is clear from Figure 2 that good and bad predictions are almost uniformly distributed. The mean true and mean false percentages are 44.95 % and 42.32 %, respectively. The percentage of good predictions (23.31 %, 193 of 828 chains) is higher than the percentage of very bad predictions (13.29 %, 110 chains), which is still not good enough for general purpose. The distribution of good and bad predictions is not uniform over the chain lengths as the histogram in Figure A2 in the Appendix nicely depicts. The above described prediction method, based on the 5 fastest modes, is much more successful with shorter (and thus less voluminous) chains, than with longer ones. With chains longer than 100 residues, but shorter than 200, the prediction algorithm was not satisfying at all, because it put more predictions in the lower right corner than in upper left. However, for chains shorter than 100 residues, it put much more predictions in the upper left (good predictions), than in lower right corner, which means that 5 modes may be good for those smaller proteins.

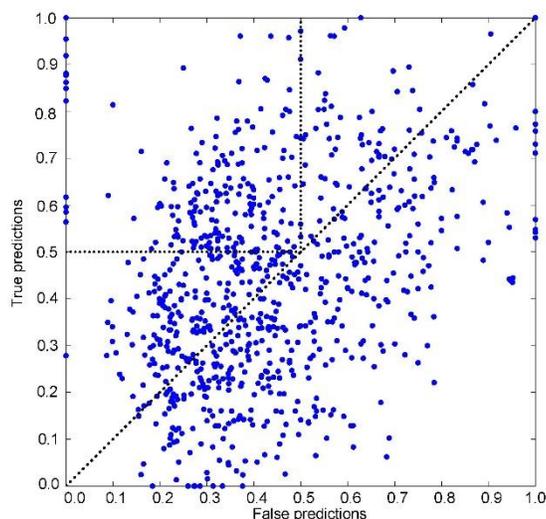

Figure 2b. The algorithm output (ratio of true vs. false predictions depicted on the 2D Cartesian space) for the simple prediction based on 5 fastest modes for each chain on our list (414 dimers in total). The diagonal line separates areas where the ratio of good predictions over the ratio of bad predictions is higher than one. The square in upper left corner is the area of good predictions, i.e. the area where more than 50 % of targets per chain is recognized as such (the ratio of true predictions vs. all targets is greater than 0.5), and the ratio of bad predictions is less or equal than 0.5 (less than 50% of non-target residues is falsely predicted). The true positives mean is 44.95 %, and the false positives mean is 42.32 %. There is 23.31 % of good predictions (193 chains, they are in upper left quadrant) and 13.29 % of very bad predictions (110 chains, they are in the lower right quadrant, which is not depicted as rectangle to emphasize the importance of good predictions).

To check whether heterodimers behave differently from homodimers, we applied the above described method on heterodimers only. Figure 3a depicts the results of that analysis. It is obvious that more predictions are put in the upper left corner, than in the lower right. That indicates that hot residues and their neighbors, recognized using only 5 fastest modes, are much closer to binding patches on the surface and in the interior of heterodimer chains. On average, there are 51.07 % of true positives, and 43.58 % of false positives. The distribution of good and very bad predictions is better than with the whole set, see Figure A3 in Appendix, but still not satisfactory enough, because there is only 31.25 % of good predictions and 11.03 % of bad ones. That means that prediction algorithm has to be improved.

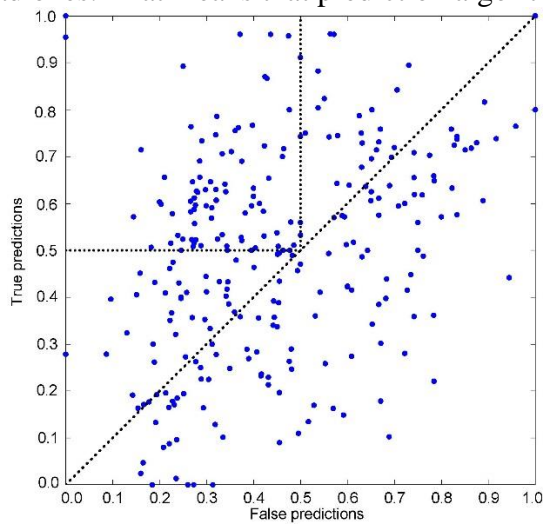

Figure 3a. The algorithm output (ratio of true predictions vs. ratio of false predictions projected on the 2D Cartesian space) for the simple prediction based on the 5 fastest modes for heterodimers only. The true positives mean is 51.07 %, and the false positives mean is 43.58 %. There is 31.25 % of good predictions (85 of 272 chains, they are in upper left quadrant) and 11.03 % of very bad predictions (30 chains, they are in the lower right quadrant).

## 2. Second approach to prediction

### *a)* Predictions based on the fastest 10% of modes

The previous analysis and the corresponding distribution of good and bad predictions over the chain lengths indicate that five fastest modes are not an ideal tool to decipher the contact patterns on the surface and in the interior of proteins. With shorter chains (up to 100 residues) they produce, via the weighted sum, a good prediction of binding and FLR residues (see Figures A2 and A3 in the Appendix). With longer chains, their prediction efficiency fails, they require more modes. We assumed that the number of fast modes used in the weighted sum calculation (Eq. 3) should be adapted to each individual protein chain. However, it would be difficult to determine the number of modes knowing only the length of the protein chain. For example, the top ten percent of modes may correspond better to the hot residues that determine protein-protein interaction, and that number is different in each individual case. Those ten percent depend on the length of a given amino acid chain, but on its three dimensional configuration also. The number of the modes covering top 10% of eigenvalues is not a linear function of the chain length. Figure 4 nicely depicts that fact. The top ten percent of eigenvalues are covered by 5 modes only for the chain H from a protein with the pdb code 1ETT which is made of 231 amino acid residues. The chain P from dimer 1BVN, has 14 modes covering top 10%, and the chain A from 1QGK has 29 modes covering top 10%.

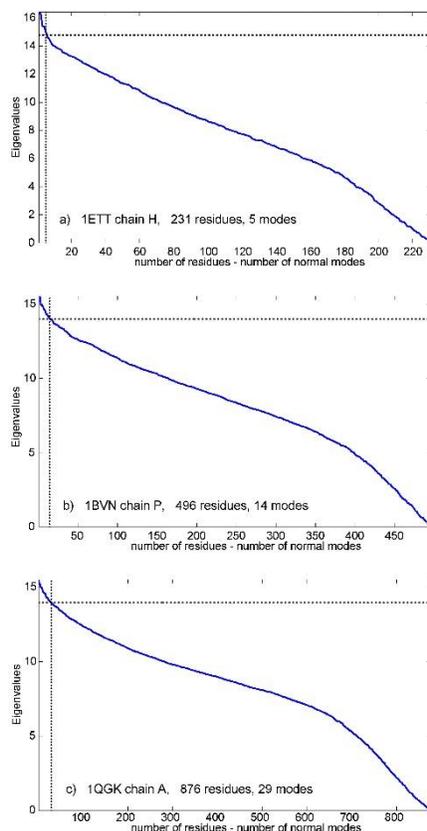

Figure 4. Distributions of eigenvalues for three different protein chains (dimer 1ETT chains H, dimer 1BVN chain P and dimer 1QGK chain A). The intersection of horizontal and vertical line on each plot designates the eigenvalues which cover to 10% of modes. It can be easily observed that top 10 % eigenvalues are covered by a different number of modes for each of these three chains. 5 modes correspond to top 10 % of eigenvalues only for 1ETT's chain H, 1BVN chain P requires 14 modes and chain A from dimer 1QGK requires 29 modes.

When this method, based on the variable number of modes, is applied together with the weighted sum calculation (Eq. 3), the percentage of true positives is increased. That is visible in Figure 5a. The same four proteins we used previously all have the number of true positives increased. However, the percentage of false positives also got increased. When this approach is applied to the heterodimers set, 136 chains in total, the improvement is miniscule, the true positives mean is 52.63 %, and the false positives mean of 47.03 %. The increase of false positives decreased the number of good predictions to 62 out of 272 chains, (22.79 % of the total number of chains), with 14.34 % of very bad predictions (39 chains), a significant increase considering only 11.03 % of false predictions with the previous approach based on the fastest 5 modes. The distribution of good vs. very bad predictions over the chain lengths again also shows that this approach is not ideal (Figure A4 in the Appendix). The bad predictions are dominant for chains longer than 100 residues, and shorter than 200 residues.

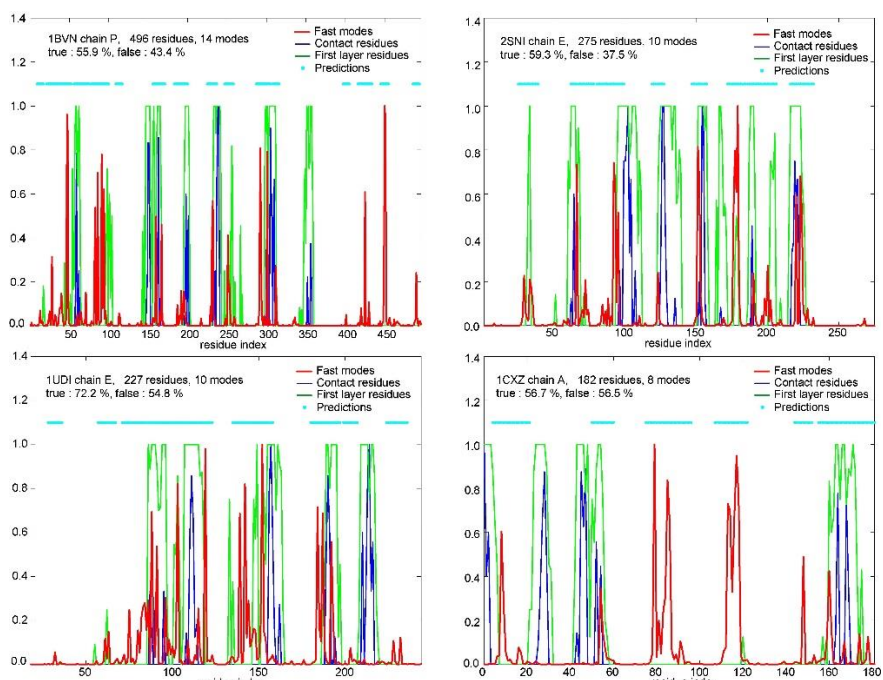

Figure 5a. The example of the simple 1D prediction (sequential neighbors influence only) based on the fastest 10 % of modes per chain, for 4 different chains (1BVN chain P, 2SNI chain E, 1UDI chain E and 1CXZ chain A). Red lines depict the weighted sums. Blue lines are contacts residues. Green lines are first layer residues. Cyan dots are predictions.

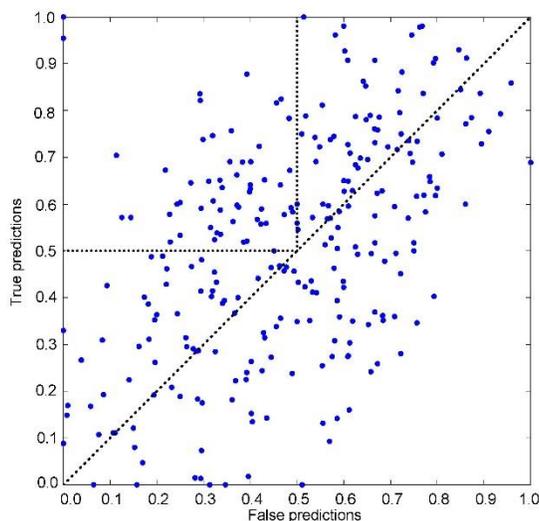

Figure 5b. The prediction output (ratio of true predictions vs. ratio of false predictions projected on the 2D Cartesian space) for the simple approach based on the fastest 10 % of modes per each chain. The true positives mean is 52.63 %, and the false positives mean is 47.03 %. There is 22.79 % of good predictions (62 of 272 chains, they are in upper left quadrant) and 14.34 % of very bad predictions (39 chains, they are in the lower right quadrant).

### *b)* **Prediction improvement - dimers with different sequence lengths**

Heterodimers are protein complexes made of two amino acid chains with no sequential and structural similarity. What kind of interaction connects different molecular entities? In protein-DNA or protein-lipid interactions, electrostatic forces are the key binding ingredients. Such forces have a small influence in protein-protein interactions.

When a protein dimer is analyzed, one may wonder whether its two constituents evolved separately, or were they created by a mutation that broke a single protein chain into two separate parts? Such mutation can easier survive if a point of separation is toward the end (or beginning) of the initial, single chain. In that case, the longer sub-chain will probably preserve its fold and function. It will also have more chances of surviving the environmental pressures. The probability of surviving is much higher than with mutations which break a protein into constituents of similar sizes. That may also mean that a longer chain produced by that single mutation, when interacting with its shorter partner (if that partner survived through evolution) may preserve its fold during (and upon) the binding. However, we can also use the reverse logic, namely that shorter constituents produced by a sequence breaking mutation preserve their fold during the binding process, only because they are shorter and easier to fold and later fit into the longer partner's pocket(s). Furthermore, if the dimer constituents evolved separately, longer partners may be less prone to significant structural changes during the binding. All this may imply that kinetically hot residues determine the binding spots in heterodimer chains. To test this assumption we separated the list of 136 heterodimers into two groups according the length ratios of their chains. We put dimers with sequence length ratios higher than 2 into a separate group. We also eliminated chains with sequence lengths smaller than 80 from this group to eliminate examples with high percentage of both true and false positives. Figure 5 depicts the analysis of the heterodimer chain lengths. Panel *a* depicts chain length for each dimer, longer chain lengths are given with the green line and shorter chains with the blue line. Panel *b* depicts corresponding chain length ratios. The vertical line separates heterodimers into two above described groups. There is 102 chains which satisfy both conditions (sequence length ration higher than 2, and chain length longer than 80).

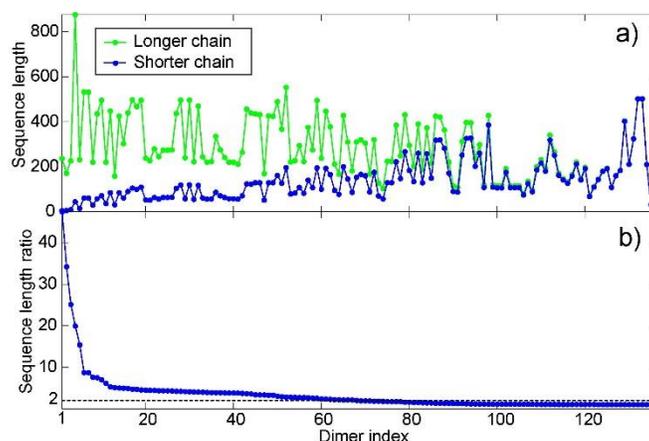

Figure 5. Dimer chain lengths for the set of 136 different heterodimers. a) Distribution of their chain lengths; longer members of a dimers are green, and the partners with shorter sequences are blue; b) Their corresponding sequence length ratios.

Figure 6 depicts the results of the analysis based on 10% of fast modes on this group of heterodimers. It is obvious that the number of proteins with badly characterized residues (proteins in which the ratio of true positives vs. false positives is less than 1) is reduced. The true positive mean is 52.28 %, and the false positive mean is 41.40 %. Although, only 6.86 % of chains belong to the group of bad predictions (7 of 102 chains), the method is still not satisfactory because only 33.33 % of all chains are in the upper left corner (34 of 102 chains, good predictions). The distribution of good vs. very bad predictions (Figure A5 in Appendix) shows much better behavior of this prediction method over the chain lengths.

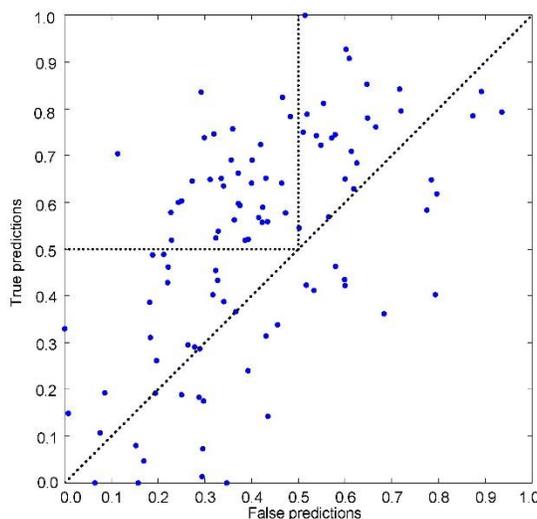

Figure 6a. Predictions output (ratio of true predictions vs. ratio of false predictions projected on the 2D Cartesian space) for the simple prediction based on the fastest 10 % of modes per chain, for heterodimer chains with high sequence length ratios (chain length ratio > 2, chain length > 80 residues). The true positives mean is 52.28 %, and the false positives mean is 41.40 %. There is 33.33 % of good predictions (34 of 102 chains), and 6.86 % of very bad predictions (7 chains).

The previous analysis performed on the heterodimer chains with low sequence length ratio and sequence length higher than 80, shows a completely different picture. There is only 12.50 % of good predictions (16 of 128 chains) versus 20.31 % of very bad predictions (26 chains). The true positives mean is 52.14 %, to the true negatives mean of 53.97 %. This proves that our assumption was correct and that chain size is important factor in protein binding. The distribution of good vs. very bad predictions (Figure A6 in Appendix) is also not very favorable to good predictions and indicates a negative correlation between kinetically hot residues and binding patches.

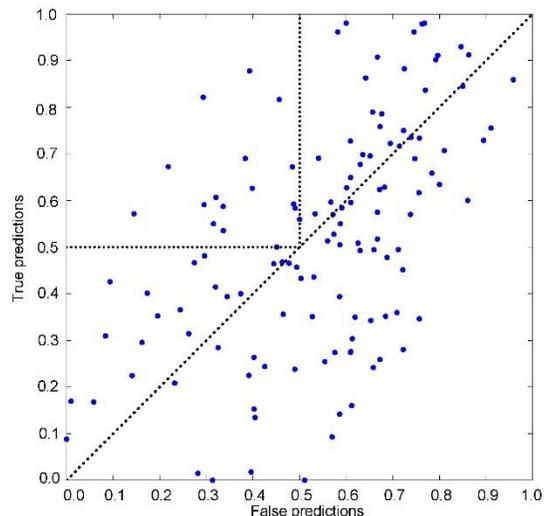

Figure 7a. Prediction output (ratio of true predictions vs. ratio of false predictions projected on the 2D Cartesian space) for the simple prediction based on the fastest 10 % of modes per chain, for chains in dimers with low sequence length ratio (Length ratio <= 2). The true positives mean is 52.14 %, and the false positives mean is 53.97 %. There is 12.50 % of good predictions (16 of 128 chains) and 20.31 % of very bad predictions (26 chains).

## 7. Prediction based on the adjustable number of modes

The previous attempts at recognizing contact and first layer residues via Gaussian network Model were based on a static approach in which protein dimer structures were analyzed using either 5 fastest normal modes, or fastest 10 % of all modes. Those approaches showed that kinetically hot residues might play a role in protein-protein interactions, but they did not offer enough proof for that. In some cases, they produced excellent results, but in other they failed. More importantly, the percentage of good predictions (the ones with more than 50% of true positives and less than 50% of false positives) was comparatively small (always less than 40 % of all samples). Many of the chains had a very high percentage of booth true positives and false positives. Also, a significant number of proteins had a very small percent of true and false predictions. The analysis of the average percent of targets per sequence length reveals that larger proteins (the ones with longer sequences) have much smaller percent of contact and first layer residues than smaller ones. Figure 8 depicts the distribution of targets over the protein sequence lengths. The figure clearly shows that small proteins have much higher ratio of contact residues than the longer ones. That information can be used to improve the prediction approach. The prediction approach should be adapted to each particular protein through the comparison of a current prediction output to the average percent of targets for that protein's sequence length class. The improvement of the prediction algorithm can be performed as follows:

− If the overall percent predictions of predictions is too large for that protein's sequence length class (for example larger than 60% of the total number of residues), the number of fast modes should be reduced by one and the whole prediction procedure should be repeated.

− If the percent of predictions is too small for the protein's sequence length class (e.g. less than 20% of all residues), the number of fast modes should be increased by one and the whole prediction procedure should be repeated.

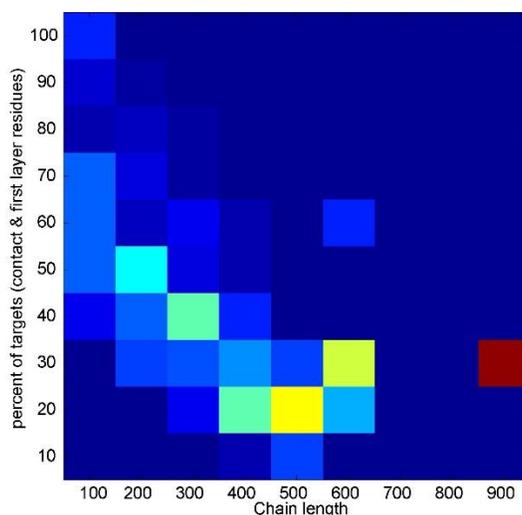

Figure 8. Distribution of targets (contact and first layer residues) per sequence length (percentage). Red color squares designate lengths (chains) with highest percentage of targets. Yellow squares are lengths with medium percentage of targets. Light blue squares are lengths with lower number of targets. Dark blue areas are chains lengths with no recognized targets at all (for our set of protein dimers).

### a) Simple, 1D linear prediction

We adopted a simple strategy to adjust the number of modes. If the number of residues in chain is less or equal than 300, it is taken than 60% or more predictions is too many. If the number of residues is greater than 300, too many predictions is taken to be more than 50% of the number of residues. Furthermore, if the chain length is less or equal to 500 residues, too few predictions is taken to be 40% of the number of residues. For longer chains, too few predictions is 20%. The prediction itself spreads the influence of kinetically hot residues linearly upstream and downstream along the sequence, as in previous cases. This approach ensures that longer chains have enough predictions, and on the other hand, the shorter ones are not saturated with not too many predictions which produce an increase of both true and false positives. Figure 9a shows how this adaptable approach works on the four examples we used previously. For the three longest chains, 1BVN chain P, 2SNI chain E, 1UDI chain E, the percent of true positives is over 50%, and percent of false positives less than 50% (1UDI, chain E has a highest difference between the two signals indicating a high correlation between the kinetically hot residues and contact patches). Only the shortest example, 1CXZ chain A, has both true and false positives over 50 %. The analysis performed on the rest of heterodimers with high sequence length ratio (length ration > 2 and chain length higher than 80) shows remarkable improvement over the previous prediction attempts. The true positives mean is 53.43 %, and the false positives is 42.34 %. There is 56.86 % of good predictions (58 of 102 chains) and only 15.69 % of very bad predictions (16 chains). The distribution of good and bad predictions is also very favorable (Figure A7 in Appendix). When this approach is applied to the heterodimers with low sequence length ratio, the true positives mean is 49.48% and the false positives mean is 49.79%. There is 32.81% of good predictions (42 of 128 chains) and 28.91% of very bad predictions (37 of 128 chains).

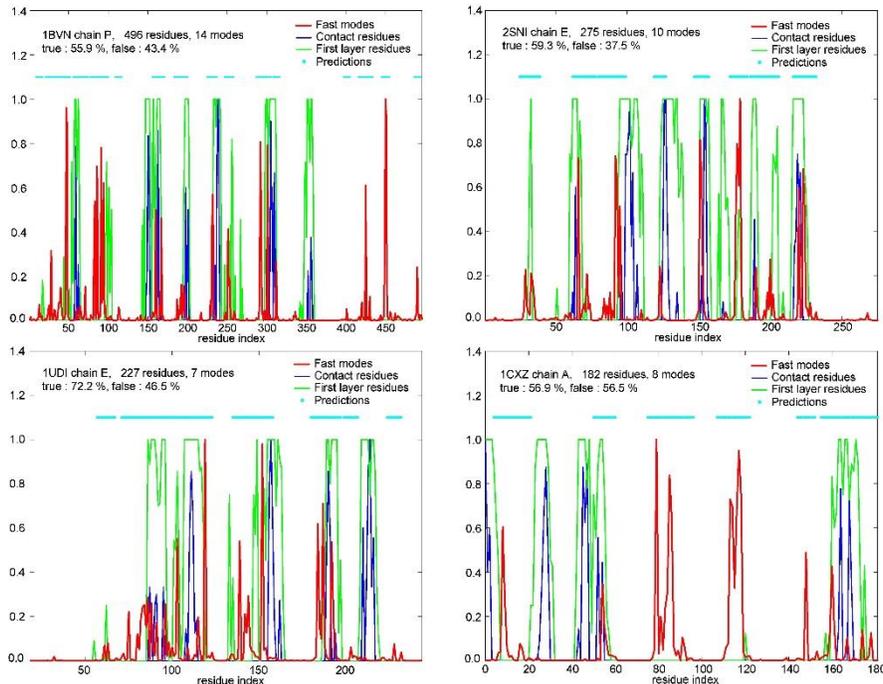

Figure 9a. An example of the prediction based on the adjustable number of fast modes and 1D influence of a hot residue (sequential neighbors influence only), for 4 different chains (1BVN chain P, 2SNI chain E, 1UDI chain E and 1CXZ chain A). Red lines depict the weighted sums. Blue lines are contacts residues. Green lines are first layer residues. Cyan dots are predictions.

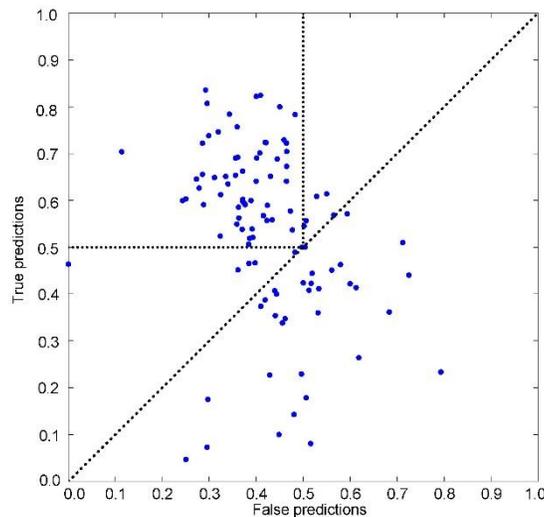

Figure 9b. Prediction output (ratio of true predictions vs. ratio of false predictions projected on the 2D Cartesian space for the prediction based on the adjustable number of fastest modes per chain and 1D influence per hot residue, for chains in dimers with high sequence length ratio (Length ratio > 2, length > 80 residues). The true positives mean true is 53.43 %, and false positives mean is 42.34 %. There is 56.86 % of good predictions (58 of 102 chains) and 15.69 % of very bad predictions (16 chains).

## 8. Going from 1D to 3D prediction

### a. Fixed influence of a hot residue

The adjustable algorithm we introduced in the previous section uses sequential neighbors to spread the influence of hot residues. It produces a good prediction of contact and first layer residues, but it offers a space for improvement (left desired). The prediction can be improved if spatial neighbors are used to spread the influence of hot residues. That would be much closer to the true nature of the GNM algorithm that uses only $C_\alpha$-$C_\alpha$ distances and disregards any sequential information. To address that, we calculated distances between residues and sorted them.

This new predictions scheme divides chains into two groups according to their sequence length. For chains shorter than 250 residues, the algorithm spreads the influence of a hot residue to its 8 closest spatial neighbors (including the hot residue), otherwise it spreads it to its 10 closest neighbors. Figure 10a depicts the algorithm output for the four examples we used previously, and Figure 10b the algorithm output for the heterodimers with the chain length longer than 80 and the sequence length ration higher than 2. The true positives mean is 52.26 %, and false positives mean is 40.39 %. There is 56.86 % of good predictions (58 of 102 chains) and only 5.88 % of very bad predictions (6 chains); see Figure A8 in the Appendix for the distribution of good and very bad predictions. There is also a noticeable number of predictions which are outside the upper left corner, and thus do not belong to the good predictions as we defined them, but with very favorable ratio of true positives vs. false positives.

When this approach is applied to the heterodimers with low sequence length ratios, the true positives mean is 49.94% and the false positives mean is 48.45%. There is 35.94% of good predictions (46 of 128 chains) and 28.91% of very bad predictions (37 of 128 chains).

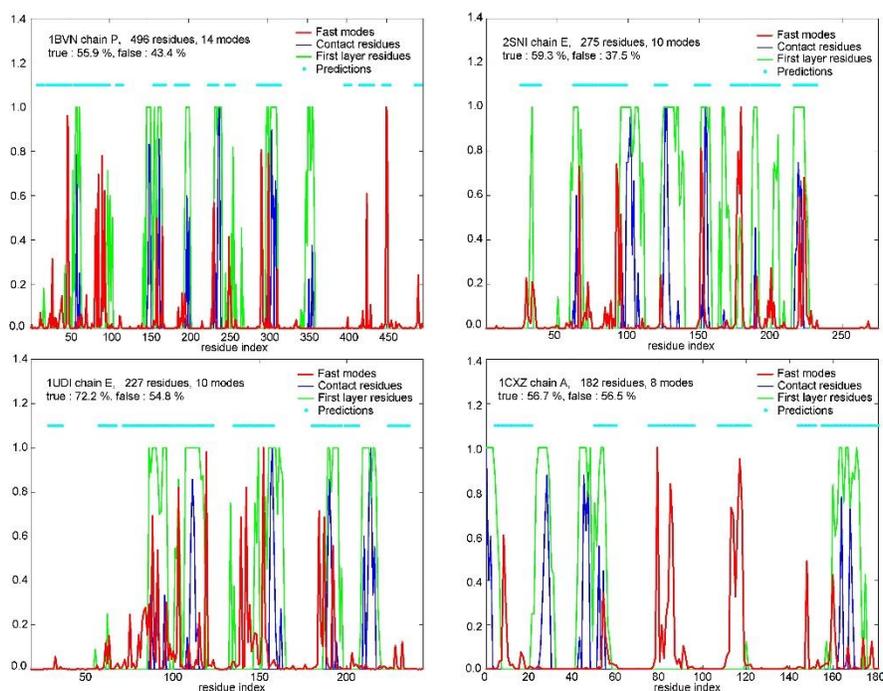

Figure 10a. An example of the prediction based on the adjustable number of fast modes and 3D influence of a hot residue (8 or 10 closest spatial neighbors), for 4 different chains (1BVN chain P, 2SNI chain E, 1UDI chain E and 1CXZ chain A). Red lines depict the weighted sums. Blue lines are contacts residues. Green lines are first layer residues. Cyan dots are predictions.

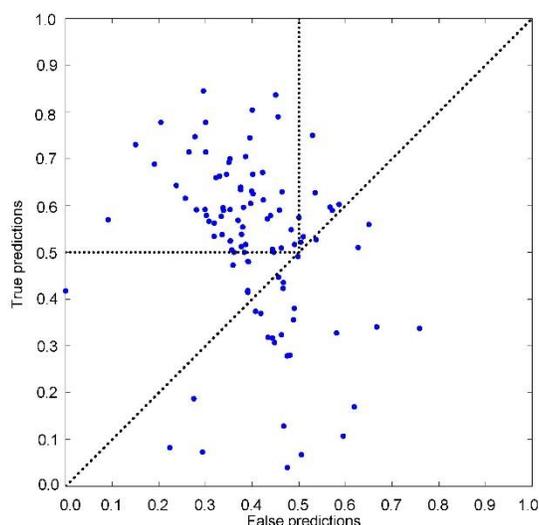

Figure 10b. Predictions output (ratio of true vs. false predictions projected on the 2D Cartesian space for the prediction based on the adjustable number of fastest modes per chain and simple 3D influence per hot residue (the influence of hot residue is spread to 8 or 10 closest residues), for chains in dimers with high sequence length ratio (Length ratio > 2, length > 80 residues). The true positives mean true is 52.26 %, and false positives mean is 40.39 %. 56.86 % of predictions are good (58 of 102 chains) and 5.88 % of very bad predictions (6 chains).

## b. Improvement of the 3D approach – variable influence per hot residue

The previous method spreads the influence of a hot residue to a fixed number of spatial neighbors, either 8 or 10 depending on the sequence length. Although aimed at recognizing contact and first layer residues it treats all residues equally, and does not take into account the differences in the density of amino acids surrounding each individual residue in the chain. The immediate neighborhood of a residue buried deep in the interior of the protein is different from the environment of residues on the surface. That information can be easily extracted from the protein structure and used with our adaptable GNM methods.

Spatial distances between $C_\alpha$ atoms of residues in a chain, when sorted and organized show variations in behavior, as Figures 11 depicts. It is obvious that in some cases, distances between a given residue and its neighbors are shorter, indicating tighter packing. In other cases, distances are much larger, meaning, that not many neighbors are in the immediate vicinity of a given residue.

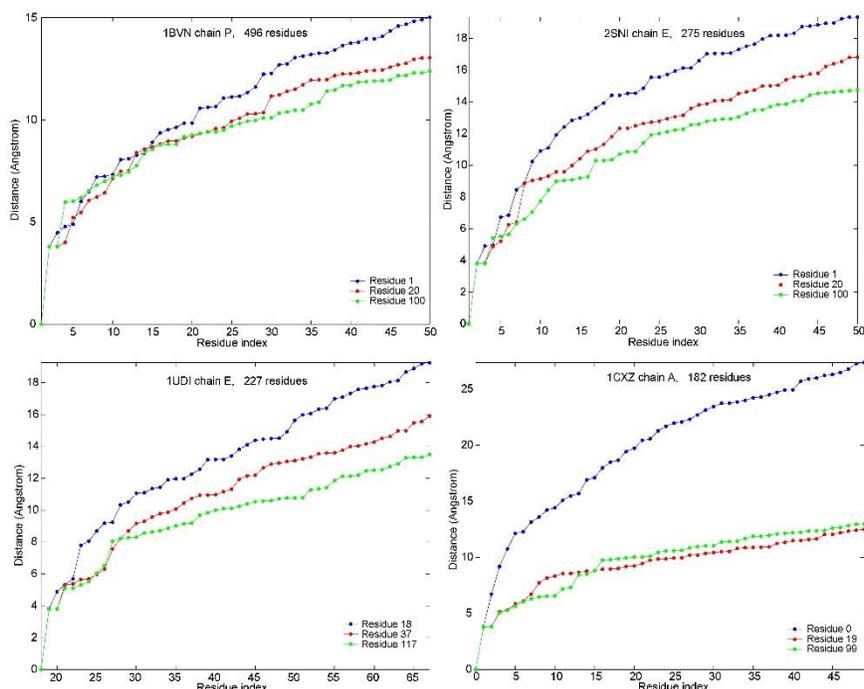

Figure 11. Spatial distances of residues in four chains (1BVN chain P, 2SNI chain E, 1UDI chain E and 1CXZ chain A) for a 3 given residue for each chain. In some cases, more than 15 residues are packed within a sphere of 10 Å, and in some cases, just a couple of residues.

We tried to use this knowledge to improve the contact and first layer residues prediction. To apply it, we first defined a maximum $C_\alpha$-$C_\alpha$ distance (a cutoff distance) to which the influence of hot residues can be spread. We used the cutoff distance of 6 Å for shorter chains (for amino-acids sequences shorter than 80) and the cutoff of 8 Å for longer chains. All residues which are within the sphere with the center in the $C_\alpha$ atom of the analyzed hot residue, and with the radius equal to the assigned cutoff distance, are considered to be "predictions", i.e., they are assumed to be either contact or first layer residues. All other residues are rejected (for that particular hot residue). To be applied, this method requires that $C_\alpha$-$C_\alpha$ distances between calculated and sorted, for each particular protein. Figure 12 depicts the results of this approach. The improvement over the previous method is minimal. The mean percentage of true positives got increased (53.51%), but the percentage of false positives got also increased by the same amount (42.12%). The distribution of good vs. bad predictions is slightly worse than in the previous case because the number of good predictions is a bit reduced (52 good predictions vs. 10 very bad predictions for 102 chains), but it is still favorable, (Figure A8 in Appendix).

When this approach is applied to the heterodimers with low sequence length ratios, the true positives mean is 51.37% and the false positives mean is 49.60%. There is 37.50% of good predictions (48 of 128 chains) and 27.34% of very bad predictions (35 chains).

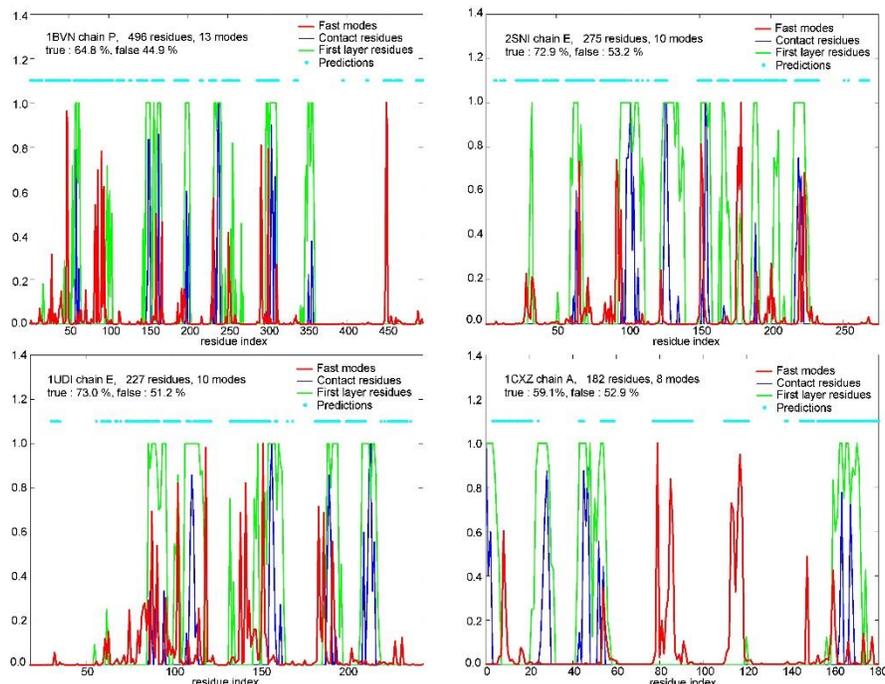

Figure 12a. An example of the prediction based on the adjustable number of fast modes and variable 3D influence of a hot residue (the influence of a hot residue is spread to spatial neighbors closer than 6 or 8 Å), for 4 different chains (1BVN chain P, 2SNI chain E, 1UDI chain E and 1CXZ chain A). Red lines depict the weighted sums. Blue lines are contacts residues. Green lines are first layer residues. Cyan dots are predictions.

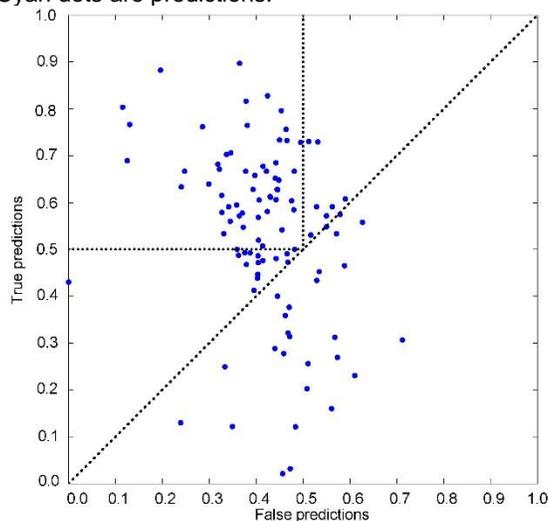

Figure 12b. Predictions output (ratio of true vs. false predictions projected on the 2D Cartesian space for the prediction based on the adjustable number of fastest modes per chain and variable 3D influence per hot residue (the influence of a hot residue is spread to spatial neighbors closer than 6 or 8 Å), for chains in dimers with high sequence length ratio (Length ratio > 2, length > 80 residues). The true positives mean is 53.51 %, and the false positives mean is 42.12 %. There is 50.98 % of good predictions (52 of 102 chains), and 9.80 % of very bad predictions (10 chains).

## 9. Combined adjustable 1D & 3D approaches

The two methods described previously base their prediction on the adjustable number of modes and the spatial influence a hot residue to its neighbors. The methods offer a good recognition of contact and first layer residues, but they disregard the fact that a protein is a chain, i.e., that amino acids are physically connected. On the other hand, the method we introduced before them, in which the influence of a hot residue is spread linearly, to sequential neighbors only, see a protein only as a set of amino acids chained together (Chapter 7). In this chapter we will attempt to combine the linear approach with two 3D spatial approaches to boost the overall prediction. By combining the 1D linear approach with the 3D ones we will include the information on the residues connectivity into a sequence blind method and thus take into account the chain-like nature of proteins (GNM calculation disregards chain connectivity and uses only physical distances between $C_\alpha$ atoms to calculate the protein connectivity matrix).

## a. 1D linear approach combined with a spatial influence of a hot residue spread to 8 or 10 closest neighbors

First, we combined the 1D prediction to the 3D adjustable method in which the influence of a hot residue is spread to 8 or 10 neighboring residues, depending on the sequence length. Therefore, in this combined approach the influence of a hot residue is first spread linearly upstream and downstream along the sequence, and then the influence is spread to hot residue's closest 8 or 10 spatial neighbors. If the percent of predictions it too high or too low, the number of fast modes is decreased or increased accordingly. The predictions cycle is repeated until the percent of predictions is within a given range for the sequence length. When applied to the set of heterodimers with a high sequence length ratio (>2), and with a chain length longer than 80 residues, this approach produces the increase in the true positives mean (56.00 %) without a significant increase in the false positives mean (43.20 %). More importantly, the combined approach puts 60.78 % of proteins in the upper left corner (62 of 102 chains belong to the group of good predictions), but keeps very bad predictions at a reasonable low 11.76 % (12 chains). Figure 12a depicts the effects of this approach on the four examples used previously, and Figure 12b depicts the effects on the whole set. The distribution of good vs. very bad predictions is also very favorable (see Figure A10 in the Appendix).

When this combined approach is applied to the heterodimers with low sequence length ratios, the true positives mean is 50.67% and the false positives mean is 50.22%. There is 34.38% of good predictions (44 of 128 chains) and 27.34% of very bad predictions (35 of 128 chains).

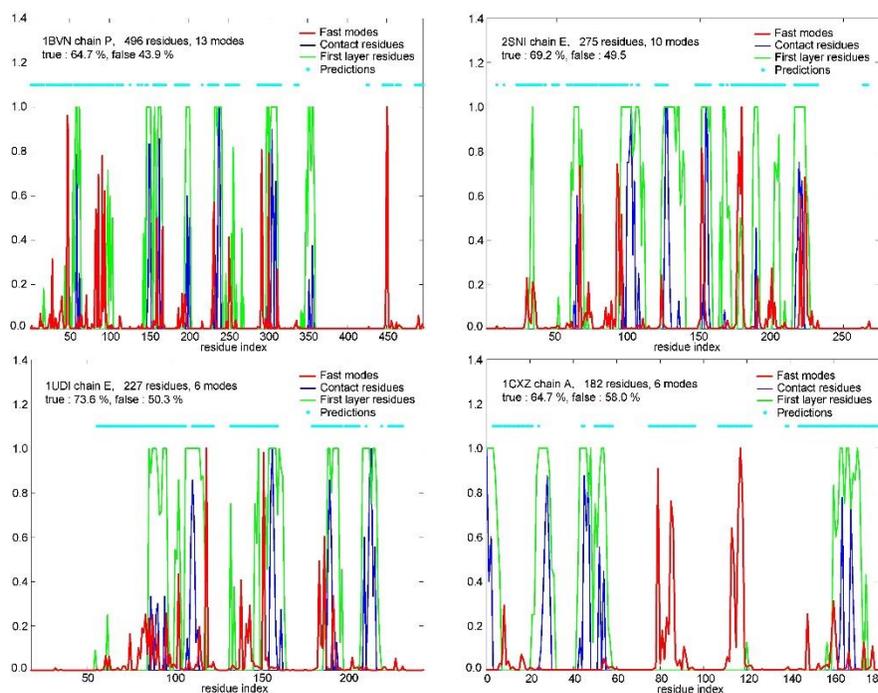

Figure 13a. An example of the prediction based on the adjustable number of fast modes and combined 1D & fixed 3D influence of a hot residue (first it is spread linearly, upstream and downstream along the sequence, and then the it is spread to residue's spatial neighbors, closest 8 or 10 neighbors), for 4 different chains (1BVN chain P, 2SNI chain E, 1UDI chain E and 1CXZ chain A). Red lines depict the weighted sums. Blue lines are contacts residues. Green lines are first layer residues. Cyan dots are predictions.

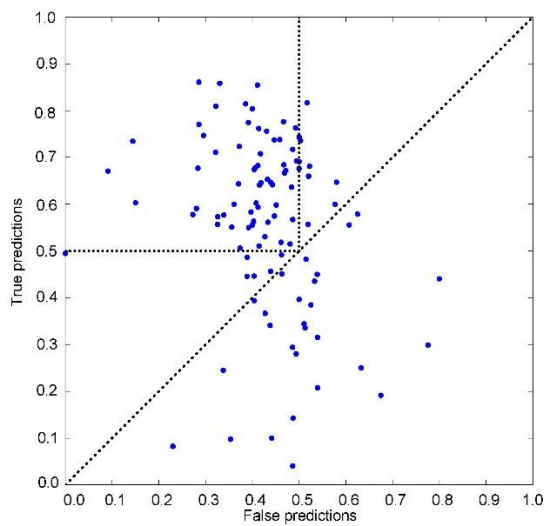

Figure 13b. Prediction output (ratio of true vs. false predictions projected on the 2D Cartesian space for the prediction based on the adjustable number of fastest modes per chain and combined 1D & fixed 3D influence of a hot residue (linearly, upstream and downstream along the sequence, and then the influence is spread to its spatial neighbors, closest 8 or 10 neighbors), for chains in dimers with high sequence length ratio (Length ratio > 2, length > 80 residues). The true positives mean is 56.00 %, and the false positives mean is 43.20 %. There is 60.78 % of good predictions (62 of 102 chains) and 11.76 % of very bad predictions (12 chains).

## b. Improvement of the combined approach

We also combined the 1D predictions with the adaptable 3D approach in which in the influence of a hot residue is variable and depends on the density of neighbors around it. We did this in attempt to boost the prediction. In this combined approach the influence of a hot residue is spread to all its neighbors, within a sphere with a given cut off radius (which may be 6 or 8 Å, depending on the sequence length). The influence of a hot residue is first spread linearly (upstream and downstream along the sequence) and then it is spread to the hot residue's spatial neighbors whose $C_\alpha$ atoms are within a sphere of a given cutoff radius with a center in the hot residue's $C_\alpha$ atom. Figure 14 (panels *a* and *b*) shows the effects of the combined approach. The differences between this and the previous 1D/liner3D combined approach are minimal and hardly noticeable. The true positives mean is 56.85 %, and the false positives mean is 43.49 %. There is 60.78 % of good predictions (62 of 102 chains) and 11.76 % of very bad predictions (12 chains). The distribution of good and bad predictions is almost the same as in the previous case. The number of good predictions for chain lengths between 200 and 300 residues is slightly reduced, but the number of good predictions for chain lengths between 300 and 500 is increased by the same amount, thus keeping the percentage of good predictions on the same level as in previous case (Figure A11 in the Appendix). This change may indicate that the method based on the variable influence of a hot residue works better with longer chains. That may be expected because longer chains have much more normal modes and offer finer resolution with the weighted sum (Eq. 3) than shorter chains.

When this combination of 1D and 3D approaches is applied to the heterodimers with low sequence length ratios, the true positives mean is 50.98% and the false positives mean is 50.51%. There is 35.16% of good predictions (45 of 128 chains) and 31.25% of very bad predictions (40 of 128 chains).

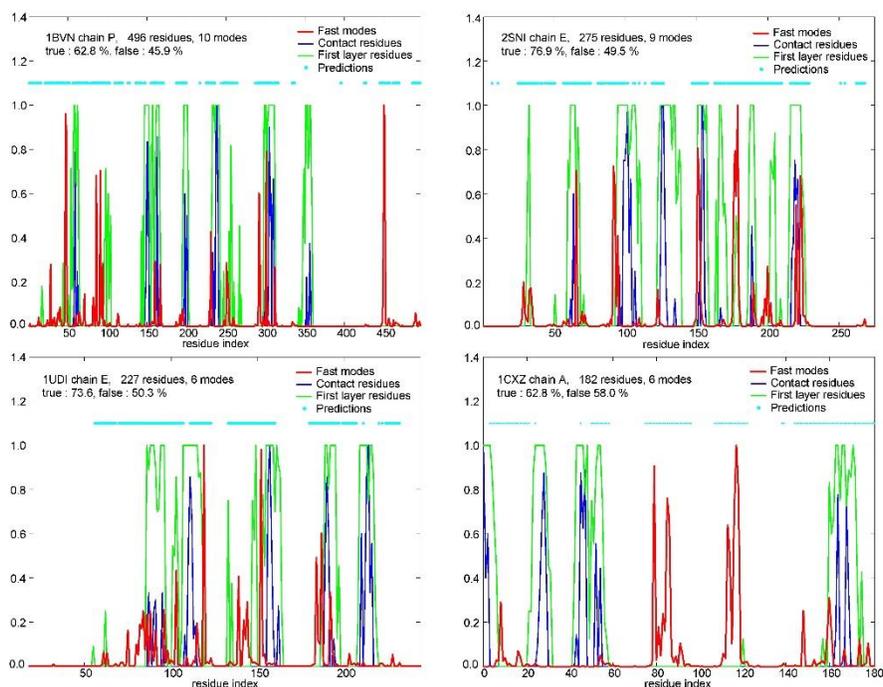

Figure 14a. An example of the prediction based on the adjustable number of fast modes and combined 1D & fixed 3D influence of a hot residue (first it is spread linearly, upstream and downstream along the sequence, and then the it is spread to residue's spatial neighbors, the ones closer than 6 or 8 Å), for 4 different chains (1BVN chain P, 2SNI chain E, 1UDI chain E and 1CXZ chain A). Red lines depict the weighted sums. Blue lines are contacts residues. Green lines are first layer residues. Cyan dots are predictions.

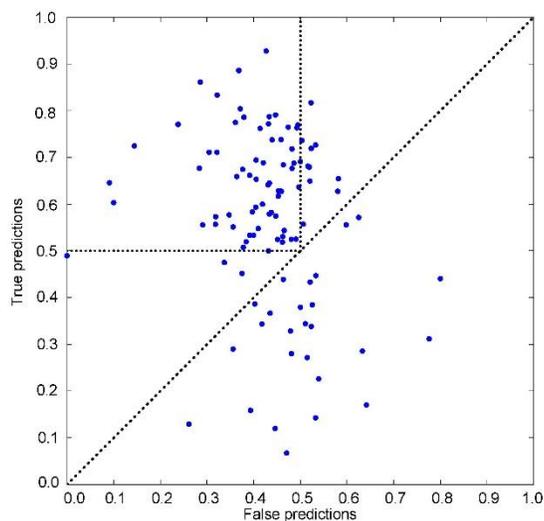

Figure 14b. Prediction output (ratio of true vs. false predictions projected on the 2D Cartesian space) for the prediction based on the adjustable number of fastest modes per chain and combined 1D & fixed 3D influence of a hot residue (first it is spread linearly, upstream and downstream along the sequence, and then the it is spread to residue's spatial neighbors, the ones closer than 6 or 8 Å), for chains in dimers with high sequence length ratio (Length ratio > 2, length > 80 residues). The true positives mean is 56.85 %, and the false positives mean is 43.49 %. There is 60.78 % of good predictions (62 of 102 chains) and 11.76 % of very bad predictions (12 chains).

## 10. Prediction algorithms comparison

In the previous sections we have presented eight methods for the contact and first layer residues prediction. We started with a very simple approach based on the fixed number of fast modes (only 5) and ended with a protocol which adjusts the number of modes to the very chain being analyzed and which spreads the influence of a hot residue along the sequence in the adjustable fashion. The true evaluation of these eight protocols can be performed only through a direct comparison of their efficiencies. The comparison of the true positives mean vs. false positives mean percentages and comparison of the percent of good vs. very bad predictions is a good measure of the prediction quality. We define a prediction to be good if more than 50% of targets of a given protein chain (contact or first layer residues) are correctly predicted, and less than 50% of all other residues are incorrectly assigned the status of contact or first layer residue (false positives).

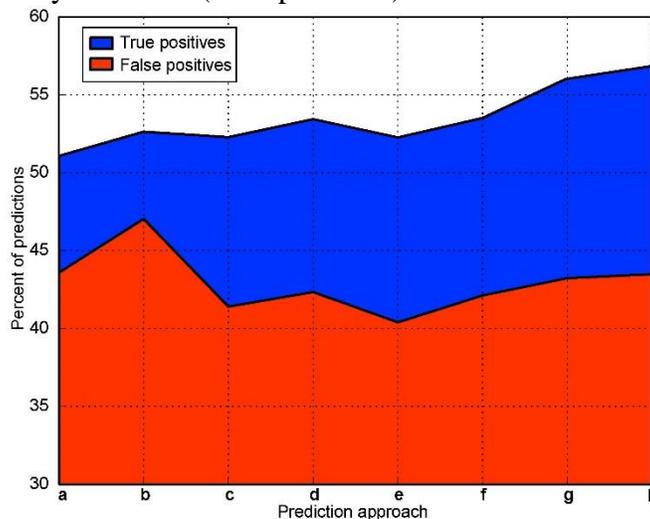

Figure 15a. Prediction algorithms comparison expressed as a plot of the true positives mean and false positives mean percentages for the each prediction algorithm described. The first two algorithms were applied to all protein chains, and the rest on chains with high sequence length ratio. The algorithms are : a) All heterodimers, 5 fastest modes; b) All heterodimers, fastest top 10% of modes; c) Heterodimers with high sequence length ratio, 10% fastest modes; d) Heterodimers with high sequence length ratio, adjustable number of modes, 1D influence; e) Adjustable modes, 3D influence, 8 or 10 closest; f) Adjustable modes, 3D influence, 6 or 8 Å; g) algorithms *d)* and *e)* combined; h) algorithms *d)* and *f)* combined.

The first three prediction approaches (Figure 15, panels a and b; five fastest modes applied to all heterodimers; fastest 10% of modes applied to all heterodimers; fastest 10% of modes applied to heterodimers with high sequence length ratio) are of a very limited ability to recognize target residues, i.e., binding patches. These methods are able to accurately predict more than 50% of target residues per chain, but they also introduce a lot of false positives. Furthermore, they are not able to put more than 40% of chains into good predictions (Figure 15b).

The first noticeable improvement is achieved when the fastest 10% of modes per chain are combined with heterodimers with high sequence length ratio (sequence length ratio higher than 2, and the number of residues in sequence higher than 80). In that case, the difference between the true positives mean and false positives mean is higher than 10 %, with the true positives mean being above 50%. However, this method is still not able to put more than 50% of protein monomers into the group of good predictions.

The true improvement is achieved only with the adjustable number of fast modes. With that approach, the difference between the mean values of true and false positives is still about 10%, but the percent of chains belonging to the group of good predictions jumps to almost 60% (Figure 15b), with less than 20% of chains belonging to bad predictions. The adjustable number of fast modes thus nicely connects kinetically hot residues to binding patches on the surface and within interior of a protein. The last four approaches, which use the 3 dimensional influence of a hot residue, are able to slightly increase the efficiency of prediction, with the last two, which combine the 1D and 3D influences, put more than 60% of proteins into the group of good predictions, while putting less than 12% of protein chains into the group of very bad predictions.

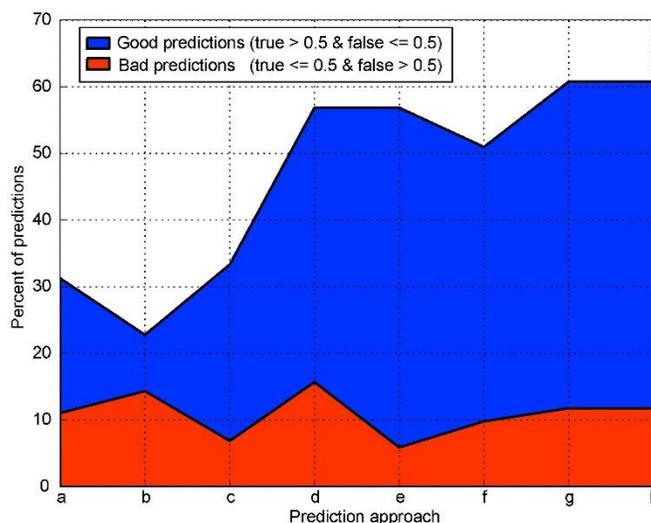
Figure 15b. Prediction algorithms comparison expressed as the percent of proteins belonging to good and very bad chain predictions for each algorithm. The first two algorithms were applied to all protein chains, and the rest on chains with high sequence length ratio. The algorithms are : a) All heterodimers, 5 fastest modes; b) All heterodimers, fastest top 10% of modes; c) High sequence length ratio, 10% fastest modes; d) Adjustable number of modes, 1D influence; e) Adjustable modes, 3D influence, 8 or 10 closest; f) Adjustable modes, 3D influence, 6 or 8 Å; g) algorithms d) and e) combined; h) algorithms d) and f) combined.

## 9. Vakser decoy sets

The Vakser decoy sets are numerically created/modeled protein structure sets (solutions) made with the intention to test protein-protein docking scoring functions. The decoy structures are based on the structures of natural protein complexes. Each decoy is a protein complex created by joining two (or more) individual single-chain structures, each based on a crystallized non-bounded protein structure. Every Vakser set contains 110 decoys with at least one near native protein complex. In most cases, each set contains 10 near native complexes.

### a. GNM algorithms and Vakser decoy sets

We tested all three adjustable algorithms (both 3D algorithms and simpler 1D approach) with the Vakser decoy sets in attempt to test the protocols efficiency in recognizing near native structures. We applied protocols only to dimer decoy sets (41 of the 72 decoy sets), because our prediction algorithms are tuned to dimer structures only. The quality of the prediction was expressed via the ratios of true positives and false positives. Those ratios can be projected on a 2D plane, which means that the standing of each decoy in every set can be expressed as its Cartesian distance from the point with coordinates (0, 1). The standing is thus, the distance of the decoy from the point that corresponds to a prediction with 0% false positives and 100% true positives (native structure).

Table 1 shows the results of our analysis. It lists all decoy sets we analyzed and the efficiency of each of the three GNM algorithms we applied. Each algorithm was applied to both chains per decoy, first to the longer chain, and then to its shorter partner. The success of each algorithm was expressed through three measures: 1) the best standing of a near native decoy (depicted through column designated as nb1 and nb2, column nb1 belongs to longer chains, and column designated nb2 to the shorter); 2) number of near native decoys among the first *n* predictions, i.e. coverage of near native decoys, where *n* is the number of near native decoys, coverage percentage, (column designated as Cov), and 3) number of near native decoys among the first *n* decoys expressed as the percentage (column designated as Cov %). The columns *f* to *k* belong to the 1D adjustable algorithm, columns *l* to *q* belong to the 3D adjustable algorithm with the fixed influence of a hot residue, and columns *r* to *w* belong to the 3D adjustable algorithm with the adaptable influence per hot residue. The last row, on the bottom of the table shows the averages for each of the columns.

| a | b | c | d | e | 1D | | | | | | 3D fixed influence | | | | | | 3D variable influence | | | | | |
|---|---|---|---|---|---|---|---|---|---|---|---|---|---|---|---|---|---|---|---|---|---|---|---|
| | | | | | f | g | h | i | j | k | l | m | n | o | p | q | r | s | t | u | v | w |
| No. | Name | sz1 | sz2 | nn | nb1 | Cov | Cov % | nb2 | Cov | Cov % | nb1 | Cov | Cov % | nb2 | Cov | Cov % | nb1 | Cov | Cov % | nb2 | Cov | Cov % |
| 1 | 1avw_A_B | 220 | 172 | 10 | 27 | 0 | 0.0% | 77 | 0 | 0.0% | 10 | 1 | 10.0% | 96 | 0 | 0.0% | 10 | 1 | 10.0% | 90 | 0 | 0.0% |
| 2 | 1bui_A_C | 247 | 121 | 10 | 78 | 0 | 0.0% | 13 | 0 | 0.0% | 8 | 1 | 10.0% | 31 | 0 | 0.0% | 9 | 2 | 20.0% | 31 | 0 | 0.0% |
| 3 | 1bui_B_C | 247 | 121 | 10 | 12 | 0 | 0.0% | 29 | 0 | 0.0% | 30 | 0 | 0.0% | 25 | 0 | 0.0% | 34 | 0 | 0.0% | 46 | 0 | 0.0% |
| 4 | 1bvn_P_T | 495 | 74 | 10 | 63 | 0 | 0.0% | 35 | 0 | 0.0% | 63 | 0 | 0.0% | 7 | 2 | 20.0% | 60 | 0 | 0.0% | 1 | 2 | 20.0% |
| 5 | 1cho_E_I | 236 | 56 | 10 | 16 | 0 | 0.0% | 3 | 1 | 10.0% | 2 | 6 | 60.0% | 8 | 2 | 20.0% | 2 | 6 | 60.0% | 6 | 3 | 30.0% |
| 6 | 1dfj_I_E | 456 | 123 | 9 | 32 | 0 | 0.0% | 71 | 0 | 0.0% | 33 | 0 | 0.0% | 63 | 0 | 0.0% | 26 | 0 | 0.0% | 67 | 0 | 0.0% |
| 7 | 1e96_A_B | 192 | 181 | 10 | 59 | 0 | 0.0% | 68 | 0 | 0.0% | 69 | 0 | 0.0% | 9 | 1 | 10.0% | 65 | 0 | 0.0% | 26 | 0 | 0.0% |
| 8 | 1ewy_A_C | 295 | 98 | 10 | 17 | 0 | 0.0% | 95 | 0 | 0.0% | 1 | 3 | 30.0% | 19 | 0 | 0.0% | 6 | 3 | 30.0% | 23 | 0 | 0.0% |
| 9 | 1f6m_A_C | 316 | 108 | 10 | 58 | 0 | 0.0% | 15 | 0 | 0.0% | 53 | 0 | 0.0% | 26 | 0 | 0.0% | 71 | 0 | 0.0% | 25 | 0 | 0.0% |
| 10 | 1fm9_A_D | 272 | 212 | 10 | 29 | 0 | 0.0% | 38 | 0 | 0.0% | 11 | 0 | 0.0% | 43 | 0 | 0.0% | 9 | 2 | 20.0% | 40 | 0 | 0.0% |
| 11 | 1g6v_A_K | 259 | 126 | 6 | 99 | 0 | 0.0% | 38 | 0 | 0.0% | 97 | 0 | 0.0% | 73 | 0 | 0.0% | 98 | 0 | 0.0% | 94 | 0 | 0.0% |
| 12 | 1gpq_A_D | 129 | 128 | 10 | 7 | 1 | 10.0% | 26 | 1 | 10.0% | 20 | 0 | 0.0% | 25 | 0 | 0.0% | 18 | 0 | 0.0% | 2 | 1 | 10.0% |
| 13 | 1gpw_A_B | 253 | 200 | 10 | 39 | 0 | 0.0% | 4 | 1 | 10.0% | 47 | 0 | 0.0% | 3 | 1 | 10.0% | 53 | 0 | 0.0% | 11 | 0 | 0.0% |
| 14 | 1he1_A_C | 181 | 131 | 10 | 46 | 0 | 0.0% | 11 | 0 | 0.0% | 39 | 0 | 0.0% | 9 | 1 | 10.0% | 45 | 0 | 0.0% | 10 | 1 | 10.0% |
| 15 | 1he8_A_B | 841 | 166 | 1 | 7 | 0 | 0.0% | 101 | 0 | 0.0% | 18 | 0 | 0.0% | 100 | 0 | 0.0% | 14 | 0 | 0.0% | 100 | 0 | 0.0% |
| 16 | 1ku6_A_B | 535 | 61 | 10 | 34 | 0 | 0.0% | 88 | 0 | 0.0% | 28 | 0 | 0.0% | 93 | 0 | 0.0% | 26 | 0 | 0.0% | 91 | 0 | 0.0% |
| 17 | 1ma9_A_B | 455 | 360 | 10 | 11 | 0 | 0.0% | 9 | 1 | 10.0% | 26 | 0 | 0.0% | 5 | 4 | 40.0% | 26 | 0 | 0.0% | 4 | 5 | 50.0% |
| 18 | 1nbf_A_D | 323 | 70 | 10 | 98 | 0 | 0.0% | 62 | 0 | 0.0% | 84 | 0 | 0.0% | 43 | 0 | 0.0% | 73 | 0 | 0.0% | 42 | 0 | 0.0% |
| 19 | 1oph_A_B | 372 | 220 | 10 | 75 | 0 | 0.0% | 1 | 9 | 90.0% | 3 | 4 | 40.0% | 1 | 5 | 50.0% | 3 | 5 | 50.0% | 1 | 5 | 50.0% |
| 20 | 1ppf_E_I | 210 | 56 | 10 | 1 | 6 | 60.0% | 2 | 3 | 30.0% | 1 | 8 | 80.0% | 32 | 0 | 0.0% | 1 | 10 | 100.0% | 52 | 0 | 0.0% |
| 21 | 1r0r_E_I | 274 | 51 | 10 | 19 | 0 | 0.0% | 48 | 0 | 0.0% | 8 | 1 | 10.0% | 10 | 1 | 10.0% | 14 | 0 | 0.0% | 21 | 0 | 0.0% |
| 22 | 1s6v_A_B | 291 | 108 | 4 | 8 | 0 | 0.0% | 10 | 0 | 0.0% | 25 | 0 | 0.0% | 13 | 0 | 0.0% | 10 | 0 | 0.0% | 10 | 0 | 0.0% |
| 23 | 1t6g_A_C | 362 | 182 | 10 | 26 | 0 | 0.0% | 4 | 2 | 20.0% | 31 | 0 | 0.0% | 42 | 0 | 0.0% | 40 | 0 | 0.0% | 33 | 0 | 0.0% |
| 24 | 1tmq_A_B | 470 | 117 | 10 | 34 | 0 | 0.0% | 84 | 0 | 0.0% | 46 | 0 | 0.0% | 96 | 0 | 0.0% | 48 | 0 | 0.0% | 97 | 0 | 0.0% |
| 25 | 1tx6_A_I | 220 | 120 | 10 | 71 | 0 | 0.0% | 59 | 0 | 0.0% | 56 | 0 | 0.0% | 57 | 0 | 0.0% | 55 | 0 | 0.0% | 83 | 0 | 0.0% |
| 26 | 1u7f_A_B | 190 | 178 | 10 | 37 | 0 | 0.0% | 44 | 0 | 0.0% | 38 | 0 | 0.0% | 41 | 0 | 0.0% | 42 | 0 | 0.0% | 41 | 0 | 0.0% |
| 27 | 1ugh_E_I | 223 | 83 | 10 | 17 | 0 | 0.0% | 8 | 2 | 20.0% | 11 | 0 | 0.0% | 5 | 2 | 20.0% | 9 | 2 | 20.0% | 5 | 1 | 10.0% |
| 28 | 1w1i_A_F | 728 | 349 | 4 | 100 | 0 | 0.0% | 11 | 0 | 0.0% | 56 | 0 | 0.0% | 10 | 0 | 0.0% | 58 | 0 | 0.0% | 15 | 0 | 0.0% |
| 29 | 1wq1_R_G | 324 | 166 | 10 | 2 | 2 | 20.0% | 72 | 0 | 0.0% | 2 | 2 | 20.0% | 67 | 0 | 0.0% | 4 | 2 | 20.0% | 68 | 0 | 0.0% |
| 30 | 1xd3_A_B | 206 | 70 | 10 | 37 | 0 | 0.0% | 77 | 0 | 0.0% | 26 | 0 | 0.0% | 73 | 0 | 0.0% | 36 | 0 | 0.0% | 74 | 0 | 0.0% |
| 31 | 1yvb_A_I | 241 | 108 | 10 | 10 | 1 | 10.0% | 7 | 2 | 20.0% | 12 | 0 | 0.0% | 2 | 6 | 60.0% | 9 | 1 | 10.0% | 14 | 0 | 0.0% |
| 32 | 2a5t_A_B | 281 | 278 | 1 | 101 | 0 | 0.0% | 54 | 0 | 0.0% | 85 | 0 | 0.0% | 58 | 0 | 0.0% | 101 | 0 | 0.0% | 96 | 0 | 0.0% |
| 33 | 2bkr_A_B | 210 | 74 | 10 | 79 | 0 | 0.0% | 10 | 1 | 10.0% | 86 | 0 | 0.0% | 5 | 2 | 20.0% | 86 | 0 | 0.0% | 6 | 1 | 10.0% |
| 34 | 2btf_A_P | 364 | 139 | 10 | 11 | 0 | 0.0% | 23 | 0 | 0.0% | 13 | 0 | 0.0% | 17 | 0 | 0.0% | 27 | 0 | 0.0% | 5 | 2 | 20.0% |
| 35 | 2ckh_A_B | 225 | 72 | 10 | 27 | 0 | 0.0% | 18 | 0 | 0.0% | 29 | 0 | 0.0% | 44 | 0 | 0.0% | 57 | 0 | 0.0% | 44 | 0 | 0.0% |
| 36 | 2fi4_E_I | 220 | 58 | 10 | 2 | 4 | 40.0% | 68 | 0 | 0.0% | 5 | 1 | 10.0% | 100 | 0 | 0.0% | 5 | 1 | 10.0% | 100 | 0 | 0.0% |
| 37 | 2goo_A_C | 103 | 92 | 10 | 101 | 0 | 0.0% | 33 | 0 | 0.0% | 101 | 0 | 0.0% | 35 | 0 | 0.0% | 100 | 0 | 0.0% | 34 | 0 | 0.0% |
| 38 | 2sni_E_I | 275 | 65 | 10 | 3 | 6 | 60.0% | 60 | 0 | 0.0% | 3 | 7 | 70.0% | 34 | 0 | 0.0% | 3 | 6 | 60.0% | 19 | 0 | 0.0% |
| 39 | 3fap_A_B | 107 | 92 | 10 | 13 | 0 | 0.0% | 27 | 0 | 0.0% | 52 | 0 | 0.0% | 25 | 0 | 0.0% | 54 | 0 | 0.0% | 27 | 0 | 0.0% |
| 40 | 3pro_A_C | 170 | 142 | 10 | 23 | 0 | 0.0% | 37 | 0 | 0.0% | 15 | 0 | 0.0% | 33 | 0 | 0.0% | 12 | 0 | 0.0% | 38 | 0 | 0.0% |
| 41 | 3sic_E_I | 275 | 108 | 10 | 4 | 5 | 50.0% | 54 | 0 | 0.0% | 1 | 6 | 60.0% | 54 | 0 | 0.0% | 2 | 5 | 50.0% | 55 | 0 | 0.0% |
| | | | | | nb1 | Cov | Cov | nb2 | Cov | Cov | nb1 | Cov | Cov | nb2 | Cov | Cov | nb1 | Cov | Cov | nb2 | Cov | Cov |
| | | | | | 37.39 | 0.61 | 6.1% | 38.88 | 0.56 | 5.6% | 32.78 | 0.98 | 9.8% | 37.37 | 0.66 | 6.6% | 34.66 | 1.12 | 11.2% | 40.17 | 0.51 | 5.1% |

Table 1. The efficiency of the 3 adjustable prediction algorithms. The columns are as follows:
a) Decoy set number,
b) Decoy set name (pdb code followed by two chain names),
c) Longer chain length,
d) Shorter chain length,
e) Number of near native structures in a set,
f) (1D adjustable algorithm) standing of the longer chain belonging to one of the best near native decoys,
g) (1D adjustable algorithm) coverage, for the longer chain, expressed as the number of near native decoys among the first *n* decoys, where *n* is the number of near native decoys (column e),
h) (1D adjustable algorithm) coverage given in column g, expressed as the percent of correctly predicted near native decoys,
i) (1D adjustable algorithm) standing of the shorter chain belonging to one of the best near native decoys,
j) (1D adjustable algorithm) coverage, for a shorter chain, expressed as the number of near native decoys among the first *n* decoy, where *n* is the number of near native decoys (column e),
k) (1D adjustable algorithm) coverage given in column j, expressed as the percent of correctly predicted near native decoys,
l) (3D adjustable algorithm) standing of the longer chain belonging to one of the best near native decoys,
m) (3D adjustable algorithm) coverage, for the longer chain, expressed as the number of near native decoys among the first *n* decoys, where *n* is the number of near native decoys (column e),
n) (3D adjustable algorithm) coverage given in column m, expressed as the percent of correctly predicted near native decoys,
o) (3D adjustable algorithm) standing of the shorter chain belonging to one of the best near native decoys,
p) (3D adjustable algorithm) coverage, for the shorter chain, expressed as the number of near native decoys among the first *n* decoys, where *n* is the number of near native decoys (column e),
q) (3D adjustable algorithm) coverage given in column p, expressed as the percent of correctly predicted near native decoys,
r) (3D adjustable algorithm with variable number of influenced residues) standing of the longer chain belonging to one of the best near native decoys,
s) (3D adjustable algorithm with variable number of influenced residues) coverage, for the longer chain, expressed as the number of near native decoys among the first *n* decoys, where *n* is the number of near native decoys (column e),

t) (3D adjustable algorithm with variable number of influenced residues) coverage given in column s, expressed as the percent of correctly predicted near native decoys,

u) (3D adjustable algorithm with variable number of influenced residues) standing of the shorter chain belonging to one of the best near native decoys,

v) (3D adjustable algorithm with variable number of influenced residues) coverage, for the shorter chain, expressed as the number of near native decoys among the first *n* decoys, where *n* is the number of near native decoys (column e),

w) (3D adjustable algorithm with variable number of influenced residues) coverage given in column v, expressed as the percent of correctly predicted near native decoys.

It is clear that the three-dimensional methods are better in recognizing near native structures than the simpler one-dimensional variant. The recognition is usually much better with a longer chain in a decoy, than with its shorter partner. The average coverage for the longer chains, for the final 3D approach with variable influence per hot residues, is twice the coverage for the shorter chains, i.e. 11.2% vs. 5.1 %. Those values are not very high, but indicate that the longer chains may be more stable during binding that their shorter partners. Figure 16 (a to d) depicts the output of the adjustable 3D GNM approach with variable number of influenced residues, as a ratio of true vs. false predictions projected on the 2D plane, for four decoys sets (1CHO, 1PPF, 1WQ1 and 2SNI). For each set, red dots depict near native structures, blue regular decoys, and green ones far from native structures. On the left panels are longer chains, and on the right panels shorter ones.

We also applied the combined 1D and 3D approaches, but found no noticeable improvement over the pure 3D predictions methods. Therefore, we did not put the analysis of the combined method here, but added it in the tabular form in the Appendix.

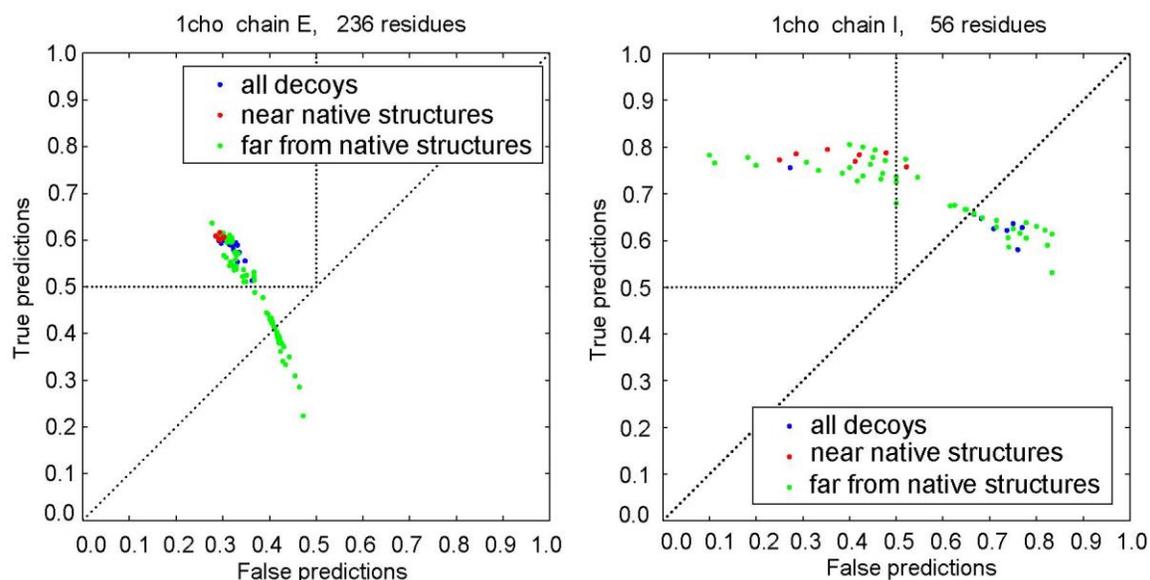

Figure 16a. An example of the GNM algorithm, with adjustable number of modes and 3D influence of a hot residue spread to spatial neighbors closer than 6 or 8 Å, applied to the decoy set 1CHO with 10 near native structures. Blue dots depict all decoys regardless their distance from a near native structure. Decoys far from native structure are colored green, and near native ones are colored red.

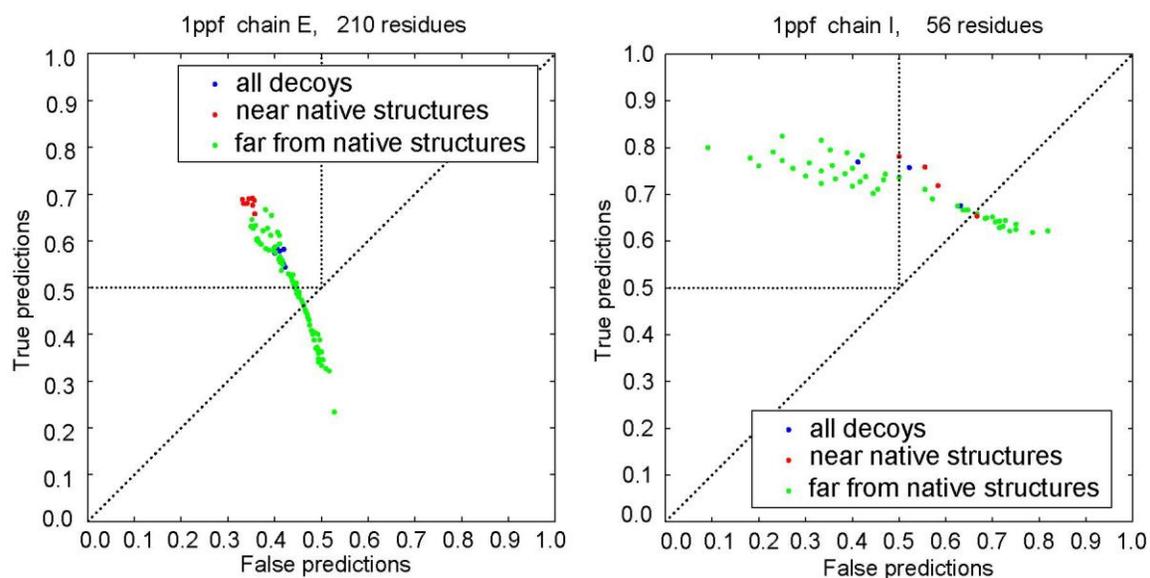

Figure 16b. An example of the GNM algorithm, with adjustable number of modes and 3D influence of a hot residue spread to spatial neighbors closer than 6 or 8 Å, applied to the decoy set 1PPF with 10 near native structures. Blue dots depict all decoys regardless their distance from a near native structure. Decoys far from native structure are colored green, and near native ones are colored red.

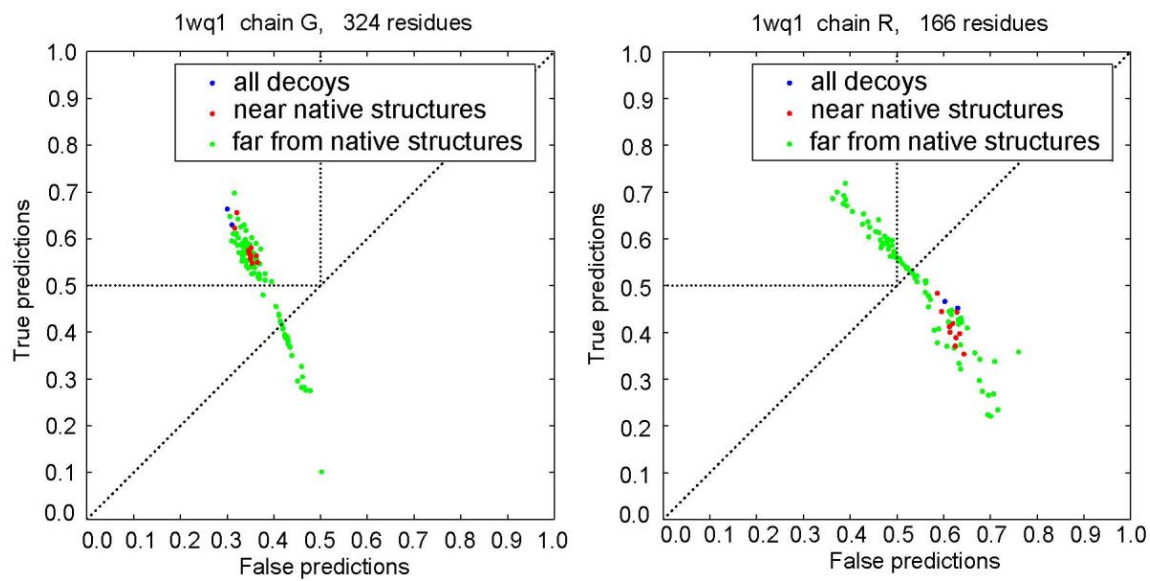

Figure 16c. An example of the GNM algorithm, with adjustable number of modes and 3D influence of a hot residue spread to spatial neighbors closer than 6 or 8 Å, applied to the decoy set 1WQ1 with 10 near native structures. Blue dots depict all decoys regardless their distance from a near native structure. Decoys far from native structure are colored green, and near native ones are colored red.

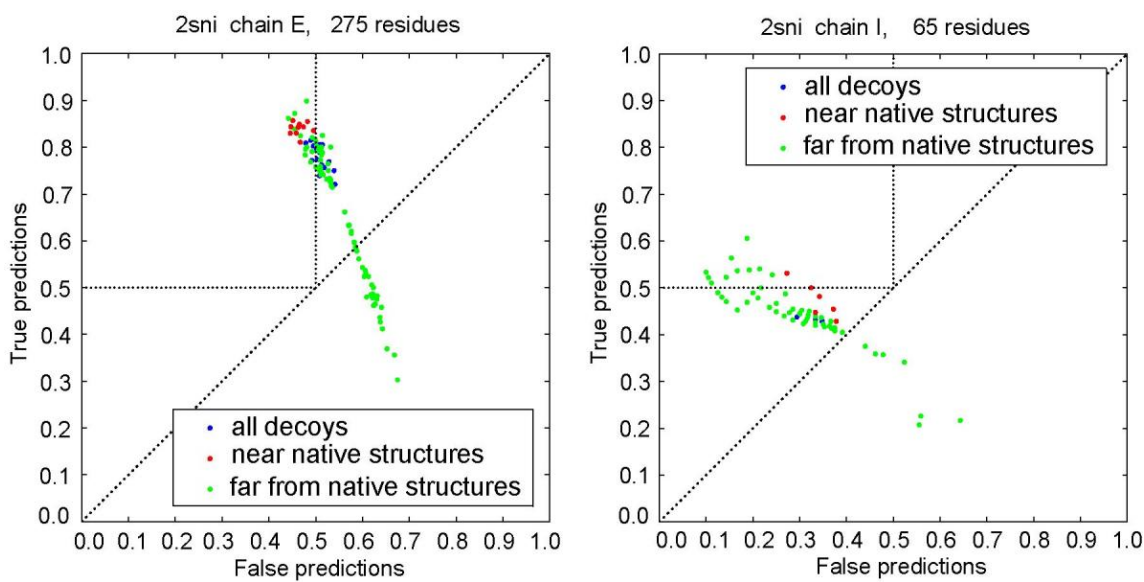

Figure 16d. An example of the GNM algorithm, with adjustable number of modes and 3D influence of a hot residue spread to spatial neighbors closer than 6 or 8 Å, applied to the decoy set 2SNI with 10 near native structures. Blue dots depict all decoys regardless their distance from a near native structure. Decoys far from native structure are colored green, and near native ones are colored red.

### b. Comparison of the statistical potential of Lu and Skolnick and adjustable GNM approach

Lu and Skolnick [2] developed a statistical method aimed at recognizing near native protein decoy structures. Their approach uses relative logarithmic frequencies of the residues appearance in protein interfaces, for all 20 types of amino acids, to calculate the interfacial energy of interacting protein chains. The relative logarithmic frequencies are contained in 20×20 matrix. The interfacial energy is calculated as a simple sum of matrix elements corresponding to the contacts residues from both chains. Thus, one energy value corresponds to each protein dimer (or multimer because statistical potential is not restricted to dimers only). The statistical potential approach produces, as its output, the binding energy for a given decoy set. That energy should be as low as possible; the lover that energy is, the better is the prediction, i.e. the more natural is the given interface. This method, however, does not produce the prediction for each individual chain in a dimer; it only gives the energy/rating for each protein complex.

We applied the statistical potential approach to every member of each of 41 decoy sets we used in the previous section and compared its ability to recognize the near native decoys to the predictions obtained with adjustable GNM prediction protocols. Table 2 shows the results of that analysis. It is obvious that the statistical potential is much better in recognizing near native decoys. Such a result can be expected because the statistical potential uses the information on residue type together with the statistical 20×20 matrix to calculate the binding energy. The GNM protocols, on the other hand, do not use the information on interacting residues nature, but apply $C_\alpha$ distances only.

| No. | Name | sz1 | sz2 | nn | Statistical potential | | |
|---|---|---|---|---|---|---|---|
| | | | | | nb | Cov | Cov |
| 1 | 1avw_A_B | 220 | 172 | 10 | 1 | 8 | 80.0% |
| 2 | 1bui_A_C | 247 | 121 | 10 | 9 | 1 | 10.0% |
| 3 | 1bui_B_C | 247 | 121 | 10 | 10 | 1 | 10.0% |
| 4 | 1bvn_P_T | 495 | 74 | 10 | 2 | 8 | 80.0% |
| 5 | 1cho_E_I | 236 | 56 | 10 | 3 | 5 | 50.0% |
| 6 | 1dfj_I_E | 456 | 123 | 9 | 1 | 6 | 66.7% |
| 7 | 1e96_B_A | 192 | 181 | 10 | 3 | 4 | 40.0% |
| 8 | 1ewy_A_C | 295 | 98 | 10 | 7 | 4 | 40.0% |
| 9 | 1f6m_A_C | 316 | 108 | 10 | 44 | 0 | 0.0% |
| 10 | 1fm9_D_A | 272 | 212 | 10 | 1 | 6 | 60.0% |
| 11 | 1g6v_A_K | 259 | 126 | 6 | 31 | 0 | 0.0% |
| 12 | 1gpq_D_A | 129 | 128 | 10 | 33 | 0 | 0.0% |
| 13 | 1gpw_A_B | 253 | 200 | 10 | 3 | 4 | 40.0% |
| 14 | 1he1_C_A | 181 | 131 | 10 | 5 | 1 | 10.0% |
| 15 | 1he8_A_B | 841 | 166 | 1 | 21 | 0 | 0.0% |
| 16 | 1ku6_A_B | 535 | 61 | 10 | 1 | 7 | 70.0% |
| 17 | 1ma9_A_B | 455 | 360 | 10 | 1 | 8 | 80.0% |
| 18 | 1nbf_A_D | 323 | 70 | 10 | 15 | 0 | 0.0% |
| 19 | 1oph_A_B | 372 | 220 | 10 | 1 | 9 | 90.0% |
| 20 | 1ppf_E_I | 210 | 56 | 10 | 7 | 1 | 10.0% |
| 21 | 1r0r_E_I | 274 | 51 | 10 | 2 | 7 | 70.0% |
| 22 | 1s6v_A_B | 291 | 108 | 4 | 2 | 1 | 25.0% |
| 23 | 1t6g_A_C | 362 | 182 | 10 | 6 | 1 | 10.0% |
| 24 | 1tmq_A_B | 470 | 117 | 10 | 1 | 6 | 60.0% |
| 25 | 1tx6_A_I | 220 | 120 | 10 | 29 | 0 | 0.0% |
| 26 | 1u7f_B_A | 190 | 178 | 10 | 14 | 0 | 0.0% |
| 27 | 1ugh_E_I | 223 | 83 | 10 | 1 | 6 | 60.0% |
| 28 | 1w1i_A_F | 728 | 349 | 4 | 4 | 1 | 25.0% |
| 29 | 1wq1_G_R | 324 | 166 | 10 | 4 | 2 | 20.0% |
| 30 | 1xd3_A_B | 206 | 70 | 10 | 1 | 10 | 100.0% |
| 31 | 1yvb_A_I | 241 | 108 | 10 | 1 | 9 | 90.0% |
| 32 | 2a5t_A_B | 281 | 278 | 1 | 11 | 0 | 0.0% |
| 33 | 2bkr_A_B | 210 | 74 | 10 | 3 | 1 | 10.0% |
| 34 | 2btf_A_P | 364 | 139 | 10 | 2 | 7 | 70.0% |
| 35 | 2ckh_A_B | 225 | 72 | 10 | 7 | 3 | 30.0% |
| 36 | 2fi4_E_I | 220 | 58 | 10 | 6 | 2 | 20.0% |
| 37 | 2goo_A_C | 103 | 92 | 10 | 13 | 0 | 0.0% |
| 38 | 2sni_E_I | 275 | 65 | 10 | 2 | 6 | 60.0% |
| 39 | 3fap_A_B | 107 | 92 | 10 | 5 | 2 | 20.0% |
| 40 | 3pro_A_C | 170 | 142 | 10 | 12 | 0 | 0.0% |
| 41 | 3sic_E_I | 275 | 108 | 10 | 1 | 8 | 80.0% |
| | Average | | | | 8.0 | 3.5 | 36.3% |

Table 2. The efficiency of the Statistical potential of Lu and Skolnick on recognizing near native structures in the Vakser decoy sets.

The ability of the statistical potential to recognize near native decoys inspired us to combine that method with the GNM adjustable methods (1D & 3D) in attempt to boost the overall prediction. We used the projection of the statuses of both GNM and statistical potential methods onto the 2D plane to score the joined method. The horizontal axis in this case corresponds to the statistical potential status of a decoy, and vertical axis to its GNM status. This allowed expression of the combined status of the two methods as a Cartesian distance from the point with coordinates 1, 1, i.e. from a point that corresponds to the best prediction according to the statistical potential and to the best prediction according to the GNM approach. Usually, none of the decoys is able to achieve that status. The results of this combined approach are shown in Table 3, which depicts the three combined approaches (statistical potential combined with 1D GNM and with both 3D GNM protocols). The first of them is the combination of 1D adjustable GNM method and statistical potential (columns $f$ to $k$). The second is the combination of the 3D adjustable GNM method and statistical potential (columns $j$ to $q$), and the third is the combination of the adjustable 3D GNM with variable number of influenced residues combined with the statistical potential (columns $r$ to $w$).

| a | b | c | d | e | GNM 1D & Statistical p.combined ||| | | | GNM 3D & Statistical p. combined ||| | | | GNM 3D adaptable & Statistical p. combined ||| | | |
| | | | | | f | g | h | i | j | k | l | m | n | o | p | q | r | s | t | u | v | w |
|---|---|---|---|---|---|---|---|---|---|---|---|---|---|---|---|---|---|---|---|---|---|---|
| No. | Name | sz1 | sz2 | nn | nb1 | Cov | Cov | nb2 | Cov | Cov | nb1 | Cov | Cov | nb2 | Cov | Cov | nb1 | Cov | Cov | nb2 | Cov | Cov |
| 1 | 1avw_A_B | 220 | 172 | 10 | 9 | 1 | 10.0% | 35 | 0 | 0.0% | 1 | 5 | 50.0% | 70 | 0 | 0.0% | 1 | 5 | 50.0% | 64 | 0 | 0.0% |
| 2 | 1bui_A_C | 247 | 121 | 10 | 44 | 0 | 0.0% | 4 | 1 | 10.0% | 2 | 3 | 30.0% | 12 | 0 | 0.0% | 2 | 3 | 30.0% | 16 | 0 | 0.0% |
| 3 | 1bui_B_C | 247 | 121 | 10 | 8 | 2 | 20.0% | 6 | 4 | 40.0% | 21 | 0 | 0.0% | 7 | 1 | 10.0% | 21 | 0 | 0.0% | 13 | 0 | 0.0% |
| 4 | 1bvn_P_T | 495 | 74 | 10 | 18 | 0 | 0.0% | 7 | 1 | 10.0% | 20 | 0 | 0.0% | 1 | 5 | 50.0% | 20 | 0 | 0.0% | 1 | 8 | 80.0% |
| 5 | 1cho_E_I | 236 | 56 | 10 | 1 | 4 | 40.0% | 1 | 8 | 80.0% | 1 | 7 | 70.0% | 2 | 6 | 60.0% | 1 | 7 | 70.0% | 1 | 5 | 50.0% |
| 6 | 1dfj_I_E | 456 | 123 | 9 | 2 | 4 | 44.4% | 34 | 0 | 0.0% | 2 | 3 | 33.3% | 23 | 0 | 0.0% | 2 | 6 | 66.7% | 31 | 0 | 0.0% |
| 7 | 1e96_B_A | 192 | 181 | 10 | 38 | 0 | 0.0% | 30 | 0 | 0.0% | 41 | 0 | 0.0% | 1 | 3 | 30.0% | 41 | 0 | 0.0% | 4 | 2 | 20.0% |
| 8 | 1ewy_A_C | 295 | 98 | 10 | 1 | 5 | 50.0% | 66 | 0 | 0.0% | 1 | 7 | 70.0% | 5 | 4 | 40.0% | 1 | 8 | 80.0% | 7 | 4 | 40.0% |
| 9 | 1f6m_A_C | 316 | 108 | 10 | 49 | 0 | 0.0% | 42 | 0 | 0.0% | 44 | 0 | 0.0% | 46 | 0 | 0.0% | 44 | 0 | 0.0% | 47 | 0 | 0.0% |
| 10 | 1fm9_D_A | 272 | 212 | 10 | 9 | 1 | 10.0% | 5 | 1 | 10.0% | 4 | 4 | 40.0% | 10 | 1 | 10.0% | 4 | 4 | 40.0% | 9 | 1 | 10.0% |
| 11 | 1g6v_A_K | 259 | 126 | 6 | 88 | 0 | 0.0% | 13 | 0 | 0.0% | 87 | 0 | 0.0% | 54 | 0 | 0.0% | 87 | 0 | 0.0% | 72 | 0 | 0.0% |
| 12 | 1gpq_D_A | 129 | 128 | 10 | 5 | 3 | 30.0% | 15 | 0 | 0.0% | 14 | 0 | 0.0% | 20 | 0 | 0.0% | 14 | 0 | 0.0% | 13 | 0 | 0.0% |
| 13 | 1gpw_A_B | 253 | 200 | 10 | 10 | 1 | 10.0% | 5 | 3 | 30.0% | 11 | 0 | 0.0% | 4 | 4 | 40.0% | 11 | 0 | 0.0% | 2 | 5 | 50.0% |
| 14 | 1he1_C_A | 181 | 131 | 10 | 19 | 0 | 0.0% | 2 | 4 | 40.0% | 11 | 0 | 0.0% | 2 | 4 | 40.0% | 11 | 0 | 0.0% | 2 | 4 | 40.0% |
| 15 | 1he8_A_B | 841 | 166 | 1 | 4 | 0 | 0.0% | 77 | 0 | 0.0% | 6 | 0 | 0.0% | 80 | 0 | 0.0% | 6 | 0 | 0.0% | 81 | 0 | 0.0% |
| 16 | 1ku6_A_B | 535 | 61 | 10 | 6 | 4 | 40.0% | 52 | 0 | 0.0% | 2 | 5 | 50.0% | 62 | 0 | 0.0% | 2 | 6 | 60.0% | 56 | 0 | 0.0% |
| 17 | 1ma9_A_B | 455 | 360 | 10 | 1 | 6 | 60.0% | 1 | 8 | 80.0% | 2 | 3 | 30.0% | 1 | 7 | 70.0% | 2 | 5 | 50.0% | 1 | 7 | 70.0% |
| 18 | 1nbf_A_D | 323 | 70 | 10 | 68 | 0 | 0.0% | 38 | 0 | 0.0% | 51 | 0 | 0.0% | 15 | 0 | 0.0% | 51 | 0 | 0.0% | 11 | 0 | 0.0% |
| 19 | 1oph_A_B | 372 | 220 | 10 | 35 | 0 | 0.0% | 1 | 10 | 100.0% | 1 | 5 | 50.0% | 1 | 9 | 90.0% | 1 | 5 | 50.0% | 1 | 10 | 100.0% |
| 20 | 1ppf_E_I | 210 | 56 | 10 | 1 | 6 | 60.0% | 1 | 8 | 80.0% | 1 | 7 | 70.0% | 2 | 4 | 40.0% | 1 | 8 | 80.0% | 6 | 2 | 20.0% |
| 21 | 1r0r_E_I | 274 | 51 | 10 | 1 | 5 | 50.0% | 16 | 0 | 0.0% | 1 | 7 | 70.0% | 2 | 1 | 10.0% | 1 | 7 | 70.0% | 5 | 1 | 10.0% |
| 22 | 1s6v_A_B | 291 | 108 | 4 | 3 | 1 | 25.0% | 2 | 1 | 25.0% | 4 | 1 | 25.0% | 3 | 2 | 50.0% | 4 | 1 | 25.0% | 2 | 2 | 50.0% |
| 23 | 1t6g_A_C | 362 | 182 | 10 | 5 | 1 | 10.0% | 12 | 0 | 0.0% | 19 | 0 | 0.0% | 37 | 0 | 0.0% | 19 | 1 | 10.0% | 32 | 0 | 0.0% |
| 24 | 1tmq_A_B | 470 | 117 | 10 | 8 | 2 | 20.0% | 44 | 0 | 0.0% | 13 | 0 | 0.0% | 62 | 0 | 0.0% | 13 | 0 | 0.0% | 65 | 0 | 0.0% |
| 25 | 1tx6_A_I | 220 | 120 | 10 | 45 | 0 | 0.0% | 35 | 0 | 0.0% | 36 | 0 | 0.0% | 44 | 0 | 0.0% | 36 | 0 | 0.0% | 71 | 0 | 0.0% |
| 26 | 1u7f_B_A | 190 | 178 | 10 | 25 | 0 | 0.0% | 20 | 0 | 0.0% | 26 | 0 | 0.0% | 18 | 0 | 0.0% | 26 | 0 | 0.0% | 18 | 0 | 0.0% |
| 27 | 1ugh_E_I | 223 | 83 | 10 | 1 | 5 | 50.0% | 1 | 6 | 60.0% | 1 | 5 | 50.0% | 1 | 5 | 50.0% | 1 | 6 | 60.0% | 1 | 4 | 40.0% |
| 28 | 1w1i_A_F | 728 | 349 | 4 | 75 | 0 | 0.0% | 1 | 4 | 100.0% | 25 | 0 | 0.0% | 1 | 4 | 100.0% | 25 | 0 | 0.0% | 1 | 4 | 100.0% |
| 29 | 1wq1_G_R | 324 | 166 | 10 | 3 | 2 | 20.0% | 47 | 0 | 0.0% | 11 | 0 | 0.0% | 36 | 0 | 0.0% | 11 | 3 | 30.0% | 36 | 0 | 0.0% |
| 30 | 1xd3_A_B | 206 | 70 | 10 | 6 | 3 | 30.0% | 52 | 0 | 0.0% | 4 | 4 | 40.0% | 45 | 0 | 0.0% | 4 | 3 | 30.0% | 47 | 0 | 0.0% |
| 31 | 1yvb_A_I | 241 | 108 | 10 | 1 | 8 | 80.0% | 1 | 5 | 50.0% | 1 | 6 | 60.0% | 1 | 9 | 90.0% | 1 | 6 | 60.0% | 1 | 8 | 80.0% |
| 32 | 2a5t_A_B | 281 | 278 | 1 | 84 | 0 | 0.0% | 23 | 0 | 0.0% | 53 | 0 | 0.0% | 28 | 0 | 0.0% | 53 | 0 | 0.0% | 73 | 0 | 0.0% |
| 33 | 2bkr_A_B | 210 | 74 | 10 | 49 | 0 | 0.0% | 2 | 5 | 50.0% | 52 | 0 | 0.0% | 1 | 6 | 60.0% | 52 | 0 | 0.0% | 1 | 7 | 70.0% |
| 34 | 2btf_A_P | 364 | 139 | 10 | 1 | 3 | 30.0% | 2 | 5 | 50.0% | 1 | 3 | 30.0% | 2 | 4 | 40.0% | 1 | 2 | 20.0% | 1 | 7 | 70.0% |
| 35 | 2ckh_A_B | 225 | 72 | 10 | 10 | 1 | 10.0% | 1 | 6 | 60.0% | 8 | 1 | 10.0% | 8 | 1 | 10.0% | 8 | 0 | 0.0% | 7 | 1 | 10.0% |
| 36 | 2fi4_E_I | 220 | 58 | 10 | 2 | 6 | 60.0% | 32 | 0 | 0.0% | 2 | 7 | 70.0% | 64 | 0 | 0.0% | 2 | 7 | 70.0% | 66 | 0 | 0.0% |
| 37 | 2goo_A_C | 103 | 92 | 10 | 86 | 0 | 0.0% | 24 | 0 | 0.0% | 85 | 0 | 0.0% | 32 | 0 | 0.0% | 85 | 0 | 0.0% | 31 | 0 | 0.0% |
| 38 | 2sni_E_I | 275 | 65 | 10 | 1 | 10 | 100.0% | 23 | 0 | 0.0% | 1 | 10 | 100.0% | 3 | 1 | 10.0% | 1 | 10 | 100.0% | 1 | 2 | 20.0% |
| 39 | 3fap_A_B | 107 | 92 | 10 | 1 | 5 | 50.0% | 12 | 0 | 0.0% | 27 | 0 | 0.0% | 9 | 2 | 20.0% | 27 | 0 | 0.0% | 16 | 0 | 0.0% |
| 40 | 3pro_A_C | 170 | 142 | 10 | 10 | 1 | 10.0% | 17 | 0 | 0.0% | 13 | 0 | 0.0% | 21 | 0 | 0.0% | 13 | 0 | 0.0% | 24 | 0 | 0.0% |
| 41 | 3sic_E_I | 275 | 108 | 10 | 1 | 10 | 100.0% | 20 | 0 | 0.0% | 1 | 10 | 100.0% | 21 | 0 | 0.0% | 1 | 10 | 100.0% | 21 | 0 | 0.0% |
| | Averages | | | | 20.3 | 2.4 | 24.9% | 20.0 | 2.0 | 21.3% | 17.2 | 2.5 | 25.6% | 20.9 | 2.0 | 22.4% | 17.2 | 2.8 | 28.1% | 23.4 | 2.0 | 22.7% |

Table 3. The efficiency of the three adjustable prediction algorithms (1D & 3D) when combined with statistical potential. The columns are as follows:
a) Decoy set number,
b) Decoy set name (pdb code followed by two chain names),
c) Longer chain length,
d) Shorter chain length,
e) Number of near native structures in a set,
f) (1D adjustable algorithm combined with statistical potential) standing of the longer chain belonging to one of the best near native decoys,
g) (1D adjustable algorithm combined with statistical potential) coverage, for the longer chain, expressed as the number of near native decoys among the first *n* decoys, where *n* is the number of near native decoys (column e),
h) (1D adjustable algorithm combined with statistical potential) coverage given in column g, expressed as the percent of correctly predicted near native decoys,
i) (1D adjustable algorithm combined with statistical potential) standing of the shorter chain belonging to one of the best near native decoys,
j) (1D adjustable algorithm combined with statistical potential) coverage, for a shorter chain, expressed as the number of near native decoys among the first *n* decoy, where *n* is the number of near native decoys (column e),
k) (1D adjustable algorithm combined with statistical potential) coverage given in column j, expressed as the percent of correctly predicted near native decoys,
l) (3D adjustable algorithm combined with statistical potential) standing of the longer chain belonging to one of the best near native decoys,
m) (3D adjustable algorithm combined with statistical potential) coverage, for the longer chain, expressed as the number of near native decoys among the first *n* decoys, where *n* is the number of near native decoys (column e),
n) (3D adjustable algorithm combined with statistical potential) coverage given in column m, expressed as the percent of correctly predicted near native decoys,
o) (3D adjustable algorithm combined with statistical potential) standing of the shorter chain belonging to one of the best near native decoys,
p) (3D adjustable algorithm combined with statistical potential) coverage, for the shorter chain, expressed as the number of near native decoys among the first *n* decoys, where *n* is the number of near native decoys (column e),

q) (3D adjustable algorithm combined with statistical potential) coverage given in column p, expressed as the percent of correctly predicted near native decoys,

r) (3D adjustable algorithm with variable number of influenced residues combined with statistical potential) standing of the longer chain belonging to one of the best near native decoys,

s) (3D adjustable algorithm with variable number of influenced residues combined with statistical potential) coverage, for the longer chain, expressed as the number of near native decoys among the first *n* decoys, where *n* is the number of near native decoys (column e),

t) (3D adjustable algorithm with variable number of influenced residues combined with statistical potential) coverage given in column s, expressed as the percent of correctly predicted near native decoys,

u) (3D adjustable algorithm with variable number of influenced residues combined with statistical potential) standing of the shorter chain belonging to one of the best near native decoys,

v) (3D adjustable algorithm with variable number of influenced residues combined with statistical potential) coverage, for the shorter chain, expressed as the number of near native decoys among the first *n* decoys, where *n* is the number of near native decoys (column e),

w) (3D adjustable algorithm with variable number of influenced residues combined with statistical potential) coverage given in column v, expressed as the percent of correctly predicted near native decoys.

The combined approach is, obviously, less successful in recognizing near native structures than the pure statistical potential. That conclusion, based on a pure statistical analysis of the prediction result (mean values), together with the statistical results of the adjustable GNM predictions may give a false impression that GNM, method based on a structural analysis, is not conclusive enough in regard the protein binding patterns. We will show that GNM, although based on a coarse-grained approach, which does not apply residue type information, produces acceptable predictions and that they are, in many instances, highly correlated with the statistical potential predictions. Figure 17 (*a* to *d*) shows some of the good predictions (the same examples were used in Figure 16). The figure depicts decoys statuses projected onto a 2D plane, as explained previously. It can be seen, especially in Fig. 17b, that statistical potential can produce significantly inferior predictions to GNM. However, the combined approach shows that contact and first layer residues prediction is again better with longer chains in decoys than with their shorter partners indicating that longer chains are more stable during the binding.

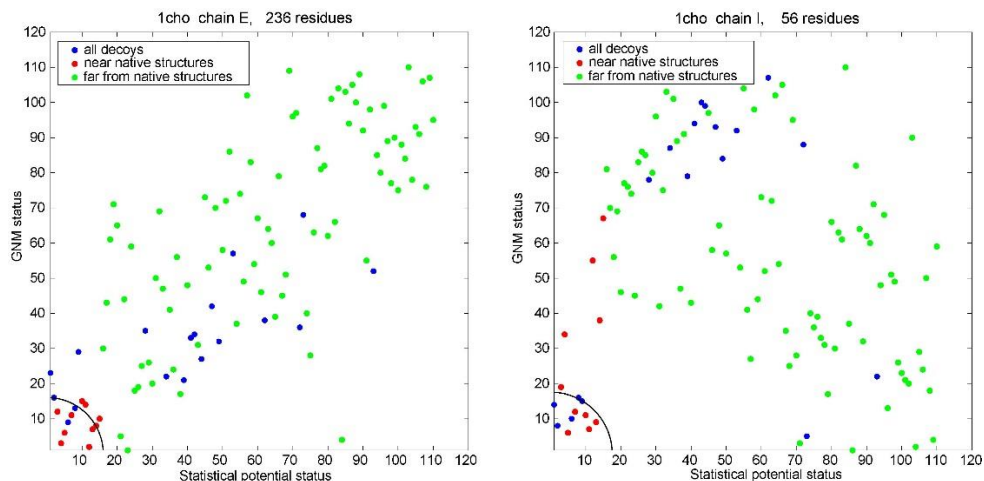

Figure 17a. An example of the GNM algorithm, with adjustable number of modes and 3D influence of a hot residue combined with statistical potential (applied to the decoy set 1CHO). Blue dots depict all decoys regardless their distance from a near native structure. Decoys far from native structure are colored green, and near native ones are colored red.

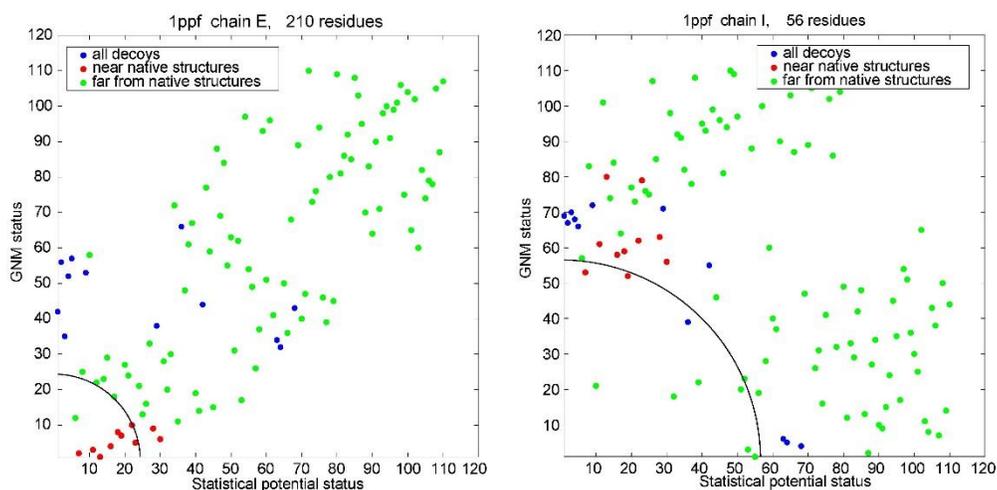

Figure 17b. An example of the GNM algorithm, with adjustable number of modes and 3D influence of a hot residue combined with statistical potential (applied to the decoy set 1PPF). Blue dots depict all decoys regardless their distance from a near native structure. Decoys far from native structure are colored green, and near native ones are colored red.

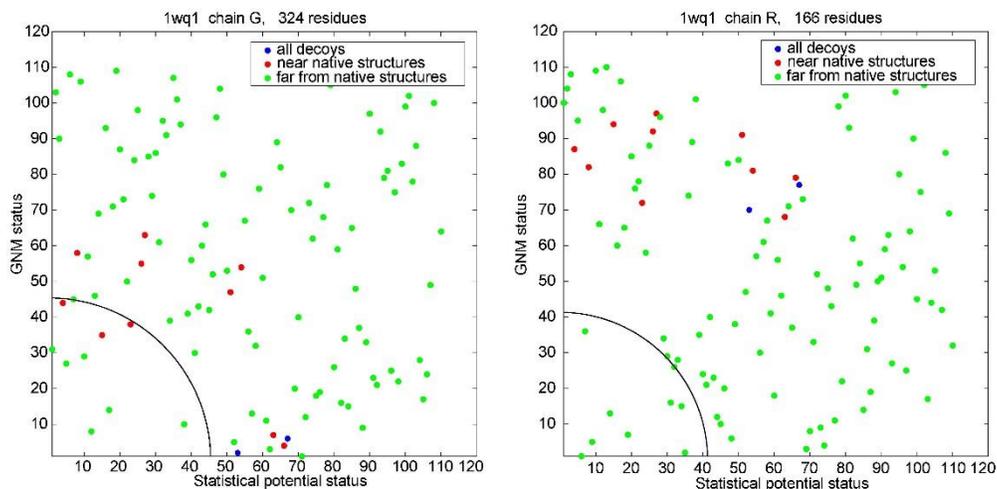

Figure 17c. An example of the GNM algorithm, with adjustable number of modes and 3D influence of a hot residue combined with statistical potential (applied to the decoy set 1WQ1). Blue dots depict all decoys regardless their distance from a near native structure. Decoys far from native structure are colored green, and near native ones are colored red.

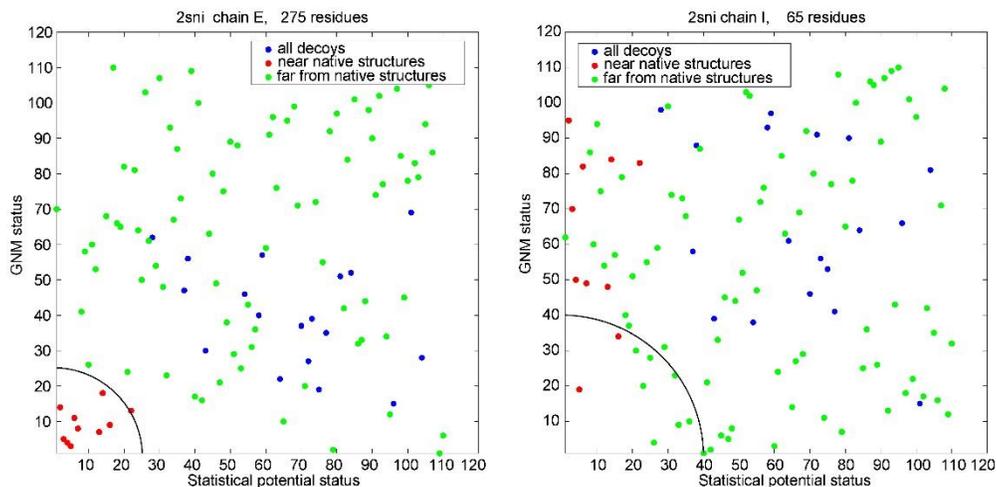

Figure 17d. An example of the GNM algorithm, with adjustable number of modes and 3D influence of a hot residue combined with statistical potential (applied to the decoy set 2SNI). Blue dots depict all decoys regardless their distance from a near native structure. Decoys far from native structure are colored green, and near native ones are colored red.

The tabular presentation of the combined approach (Table 3) is not a very clear depiction of the ability of both algorithms (GNM and statistical potential) to recognize near native decoys. We, therefore, used the stacked bars to visualize the abilities of both approaches to recognize near native decoys. Figure 18 shows the best prediction status for a near native structure for each decoy set. The upper panel depicts the status of the longer, and lower panel depicts the status of the shorter chain. The lower is the bar, the better is the prediction. The figure clearly shows that in many cases the chains behave oppositely. One chain is quite rigid (and thus its binding patterns are better described with GNM) while its partner is more flexible and thus less precisely described by the normal mode analysis.

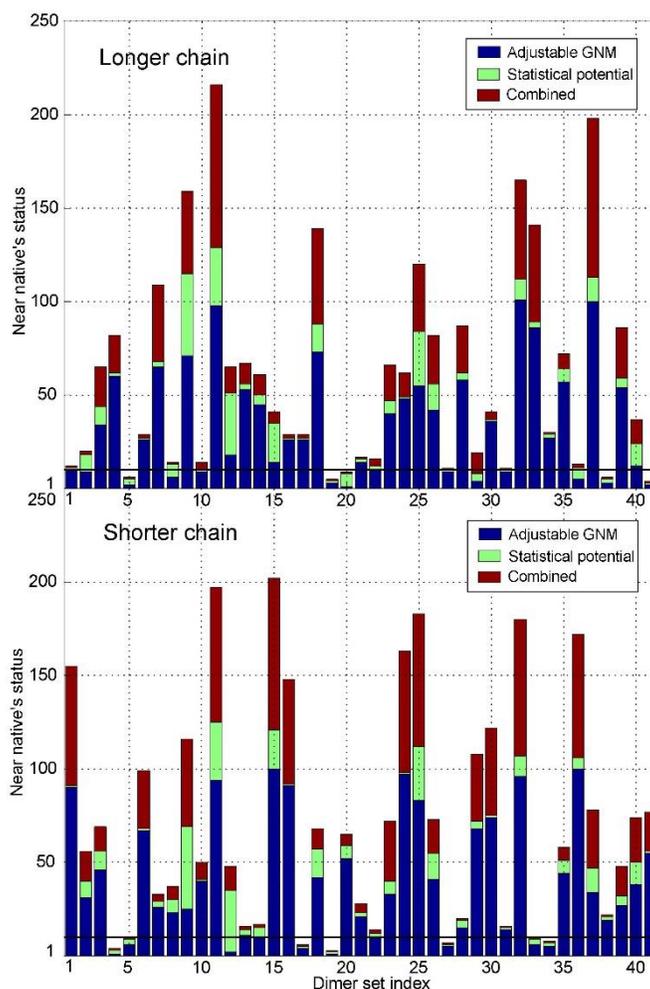

Figure 18. The status of the best near native structure as predicted with the adjustable GNM algorithm (variable 3D influence), with the Lu and Skolnick's statistical potential, and their combination. The shorter the bar, the better the prediction.

Figure 19 depicts the near native structures coverage for each decoys set. The bar height corresponds to the percentage of near native decoys among the first *n* predictions, where *n* is the number of near native decoys. The taller is the bar, the better is the prediction. That value is, thus, the coverage for the each decoy set. It is calculated as 100×(number of near native predictions among the first *n* predictions)/*n*(number of near native decoys). The difference in binding behavior between binding partners, measured by the GNM algorithm, is even more visible in this figure. In most of the cases in which the GNM algorithm is able to recognize near native decoys, it recognizes them in one chain only. The near native decoys of the other chain are mostly not recognized. That implies a different behavior during binding. One chain is rigid, and that chain is nicely recognized by the GNM algorithm, while the other chain is more flexible, and thus not well described by the GNM algorithm which emphasizes kinetically active residues with small amplitude fluctuations.

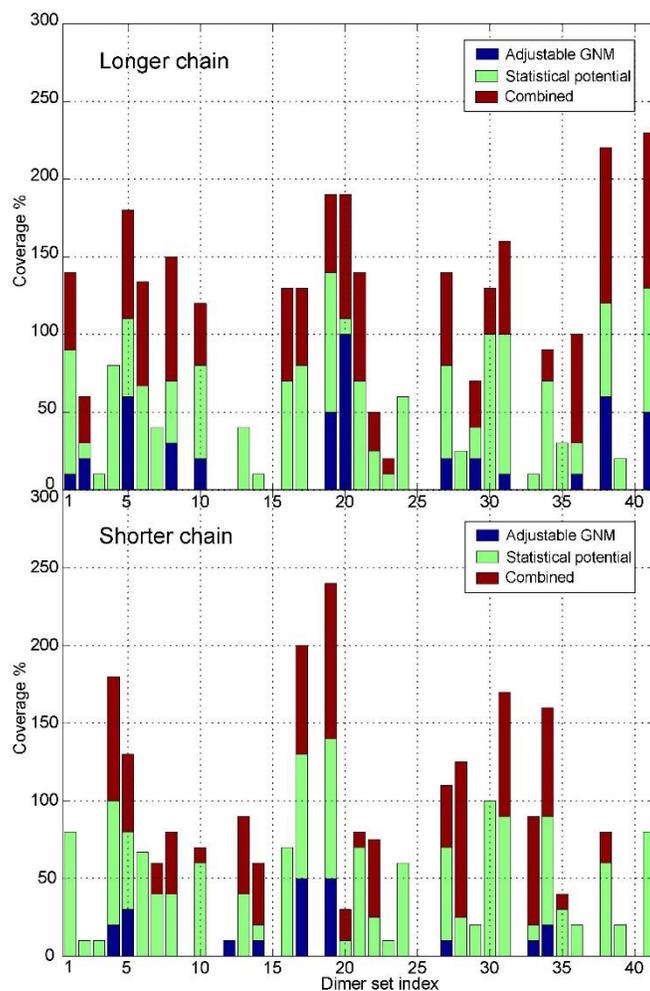

Figure 19. The percentage of near native structure's as predicted with the adjustable GNM algorithm (variable 3D influence), with the Lu and Skolnick's statistical potential, and their combination. The taller the bar, the better is the prediction.

## Conclusion

This manuscript describes a couple of algorithms for binding and first layer residues recognition we developed. The algorithms are based on the Gaussian Network Model [1] of Demirel at al. They were developed and tested on the set of 414 protein dimers (136 heterodimers and 278 homodimers). Heterodimers were early on rejected in favor of heterodimers, thus all major were applied and tested on the protein dimers made of different constituents. The pure GNM algorithm, based on 5 fastest modes per chain is not able to correctly connect kinetically hot residues and binding patches in heterodimers. The increase in the number of modes to fastest 10 % of modes does not produce the desired effect. When the number of modes used in the weighted sum calculation, is set to fastest 10%, and the GNM algorithm is applied to heterodimers only, the prediction of target residues exhibits a slight improvement. The false positives mean get decreased, and the difference in the percent of proteins with correctly recognized target residues (good predictions) and proteins with badly recognized target residues (very bad predictions) rises to more than 10%. The first noticeable improvement is achieved when the algorithm that uses fastest 10% in weighted sum calculation is applied to heterodimers with significant chain length difference of their constituents (chains longer than 80 residues, and the dimer chains length ratio higher than 2). Both false positives rate and the percent of very bad predictions (protein chains with unfavorable ratio of true positives and false positives) get decreased.

The first algorithm which can put more than 50% of chains into the group of good predictions uses the adjustable number of modes on heterodimers with significant chain length difference. The number of modes per chain is changed until the number of predictions does not correspond to the expected number of targets for the given protein. The analyzed proteins are grouped according their chain length, and for every chain length group the average percent of target residues is estimated. The true positives mean in this case stays above 50% (an average percent of recognized targets per chain is above 50%), and the false positives mean is close to 40% (an average percent of falsely assigned residues as targets per chain is about 40%). The initial adjustable approach which spread the influence of a hot residue to sequential neighbors is extended to a full 3D approach. The two 3D algorithms (the one which spread the influence of a hot residue to a fixed number of neighbors, and the one which spreads the influence according the residues spatial density) further improve the target recognition quality. The last two adjustable algorithms which combine 1D and 3D approaches (first they spread the hot residue influence to sequential neighbors, and then to spatial neighbors also) have the true positives rate of more than 55% and put more than 60% of protein monomer chains into the group of good predictions (heterodimers with high sequence length ratio, longer than 80 residues).

We also tested our algorithms on the Vakser decoy sets (41 sets of heterodimer decoys). Each decos set contains 110 decoys, of which are 10 (on average) near native. A target recognition algorithm's task is to correctly separate near native decoys from the rest of decoy constructs. The 3D adjustable algorithm which spreads the influence of a hot residue to the variable number of neighbors proved to be the most efficient. The algorithms which combine 1D and 3D approaches are not as efficient because they generate more false positives. We also compared the GNM approach to the statistical potential approach of Lu and Skolnick [2]. The statistical potential produces better overall results, but in a noticeable number of cases it is not better than the GNM approach. The advantage of the statistical potential is that it uses a full residue information, as opposite to GNM which disregards the nature of individual residues. It is interesting to out point that the decoys sets which were badly characterized by the statistical potential also did poorly with the GNM approach (see Figures 18 and 19). That means that there is more than two approaches to protein binding (kinetically stable residues, and statistically compatible residue pairs). One more impression may also be drawn from the analysis of the Vakser decoy sets. Proteins which form a heterodimer have different dynamic behavior. As can be seen in Figures 18 and 19 one of the chains forming a dimer is usually more kinetically active. Its binding patches are more easily that the binding patches of its partner (it they can be recognized at all). In much smaller number of cases binding patches on the surfaces and in the interior of both partners are characterized by the kinetically hot residues. That may imply that during binding one chain is passive and immovable, with interfacial residues which can be well characterized by GNM. On the other hand, its partner is more active, and thus has residues which cannot be well characterized by GNM.

This manuscript describes our attempt at recognizing binding behavior of proteins. Although based on a crude model which disregards the nature of interacting residues and their physical and chemical properties, the described methods are able to recognize the majority of interacting residues in at least one of the interacting partners. Our results show that at least one of the interacting chains has to have a stable binding surface.

# Appendix

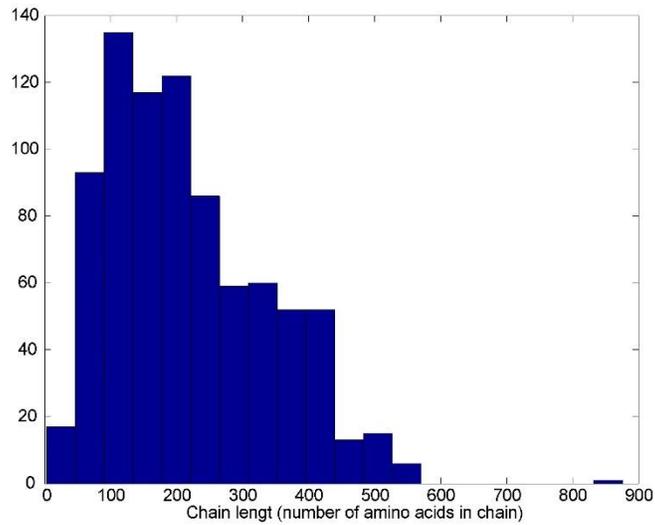

Figure A1. Protein chain lengths distribution for both heterodimers and homodimers.

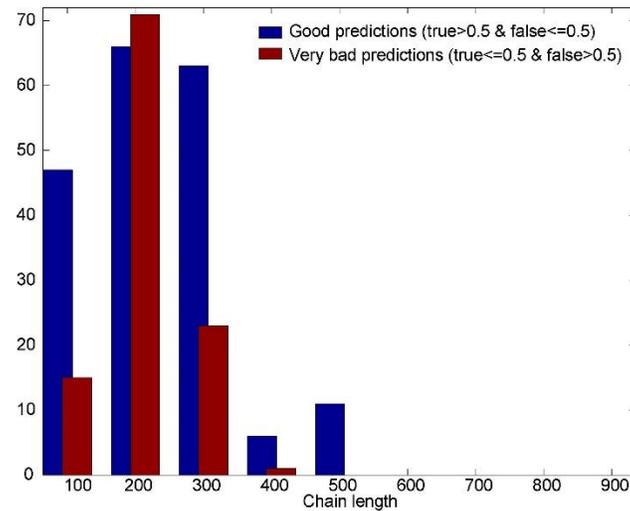

Figure A2. The prediction histogram based on the analysis of all chains (828 of them) over the sequence lengths for the simple prediction based on 5 fastest modes. Only good and very bad predictions are depicted. Blue bars are good predictions and red bars are very bad predictions. It is obvious that 5 modes do not offer good prediction because in some cases (chain longer that 100 and shorter than 200 amino acids) the number of bad predictions is higher than the number of good predictions.

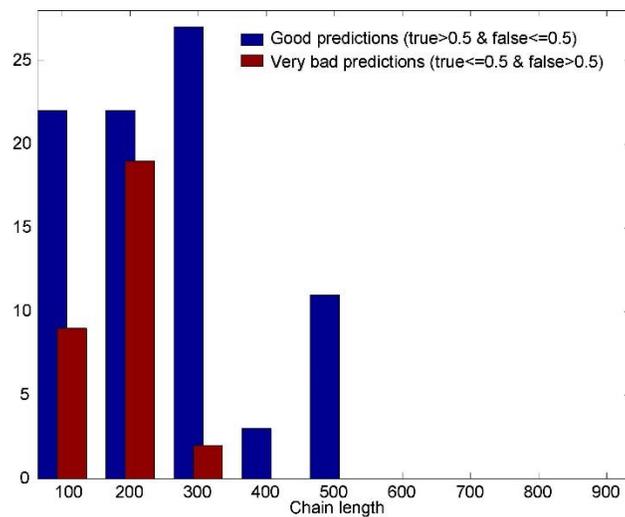
Figure A3. The prediction histogram over the sequence lengths for the simple prediction based on the 5 fastest modes for each heterodimer chain. The prediction is better than with heterodimers and homodimers combined, but not satisfactory, because there is still less than 50 % of good predictions (31.25 % of good predictions and 11.03 % of bad ones).

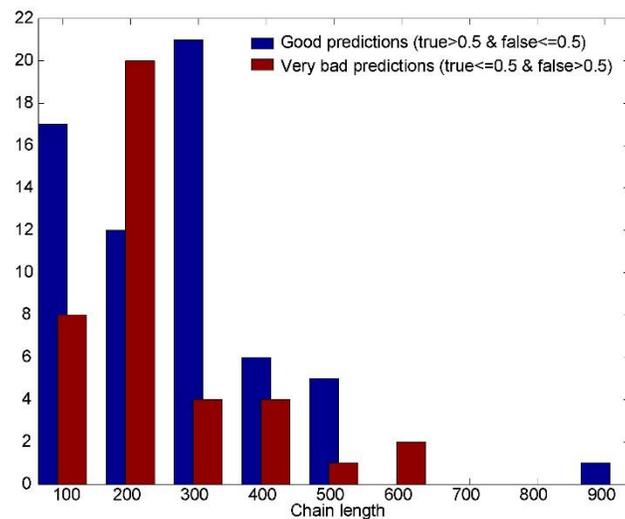
Figure A4. The prediction histogram over the sequence lengths for the simple prediction based on the fastest 10 % of modes for each heterodimer chain. There is still less than 50 % of good predictions and the distribution is worse than the distribution for 5 modes only (22.79 % of good predictions and 14.34 % of very bad predictions).

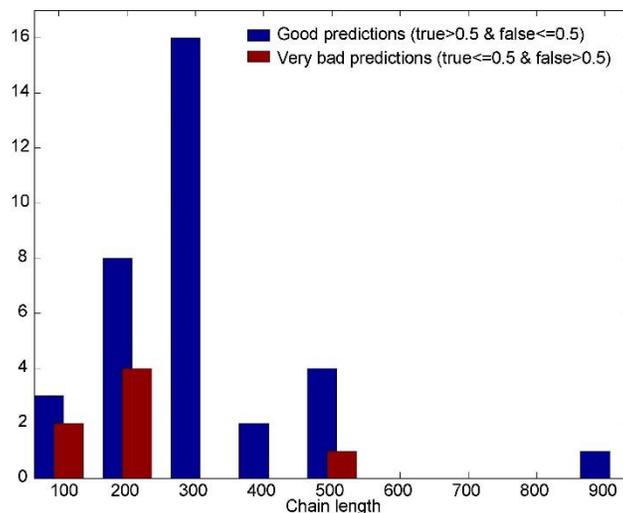
Figure A5. The prediction histogram over the sequence lengths for the simple prediction based on the fastest 10 % of modes for each chain, for chains in dimers with a high sequence length ratio (Length ratio > 2, length > 80 residues). There is still less than 50 % of good predictions and the distribution is worse than the distribution for 5 modes only (33.33 % of good predictions and 6.86 % of very bad predictions).

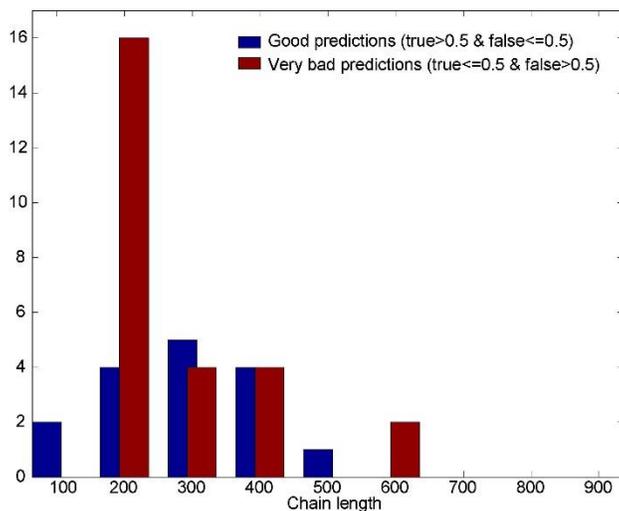
Figure A6. The prediction histogram over the sequence lengths for the simple prediction based on the fastest 10 % of modes for each chain, for chains in heterodimers with low sequence length ratio (Length ratio <= 2). Blue bars are good predictions and red bars are very bad predictions. There is only 12.50 % of good predictions to 20.31 % of very bad predictions.

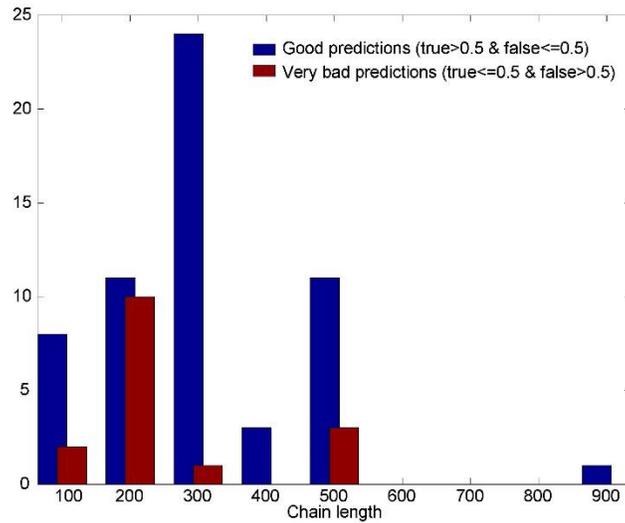

Figure A7. The prediction histogram over the sequence lengths for the prediction based on the adjustable number of fast modes, for 1D influence of hot residues, for chains in dimers with high sequence length ratio (Length ratio > 2, length > 80 residues). The true positives mean true is 53.43 %, and false positives mean is 42.34 %. There is 56.86 % of good predictions and 15.69 % of very bad predictions.

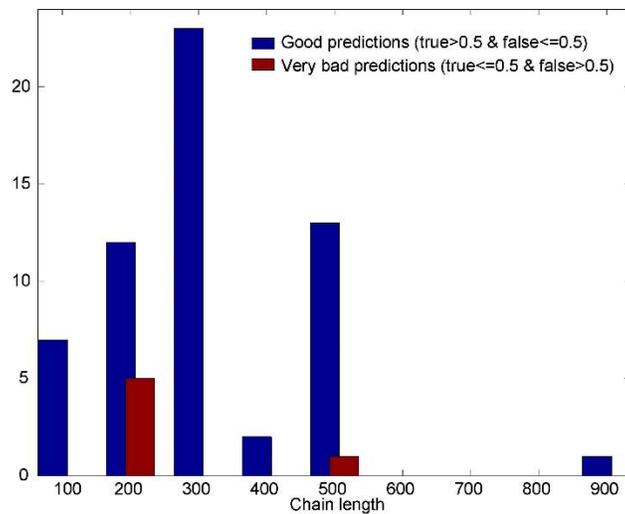

Figure A8. The prediction histogram over the sequence lengths for the prediction based on the adjustable number of fast modes and simple 3D influence of hot residues (the influence of hot residue is spread to 8 or 10 closest residues), for chains in dimers with high sequence length ratio (Length ratio > 2, length > 80 residues). The true positives mean true is 52.26 %, and false positives mean is 40.39 %. There is 56.86 % of good predictions and 5.88 % of very bad predictions.

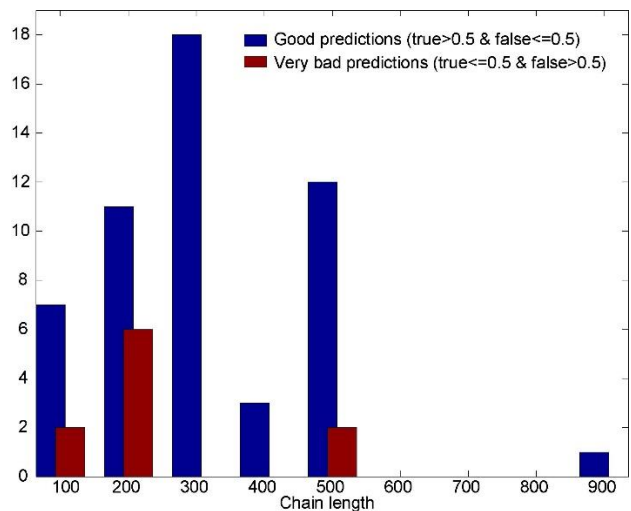

Figure A9. The prediction histogram over the sequence lengths for the prediction based on the adjustable number of fast modes and variable 3D influence per hot residue (the influence of a hot residue is spread to spatial neighbors closer than 6 or 8 Å), for chains in dimers with high sequence length ratio (Length ratio > 2, length > 80 residues). The true positives mean is 53.51 %, and the false positives mean is 42.12 %. There is 50.98 % of good predictions and 9.80 % of very bad predictions.

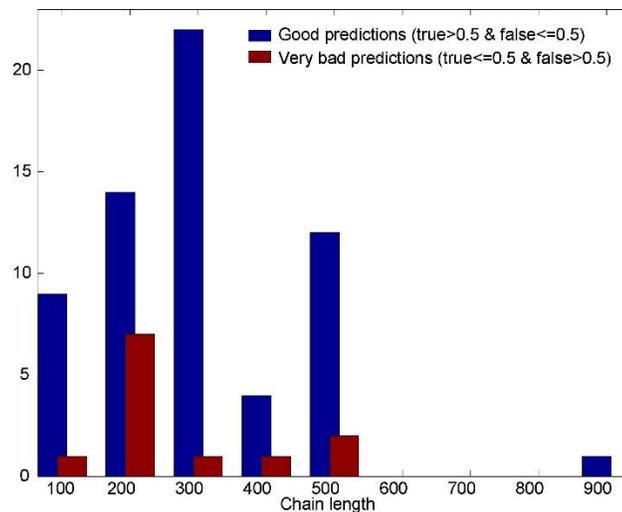

Figure A10. The prediction histogram over the sequence lengths for the prediction based on the adjustable number of fast modes and combined 1D & 3D influence of a hot residue (linearly, upstream and downstream along the sequence, and then the influence is spread to its spatial neighbors, closest 8 or 10 neighbors), for chains in dimers with high sequence length ratio (Length ratio > 2, length > 80 residues). The true positives mean is 56.00 %, and the false positives mean is 43.20 %. There is 60.78 % of good predictions and 11.76 % of very bad predictions.

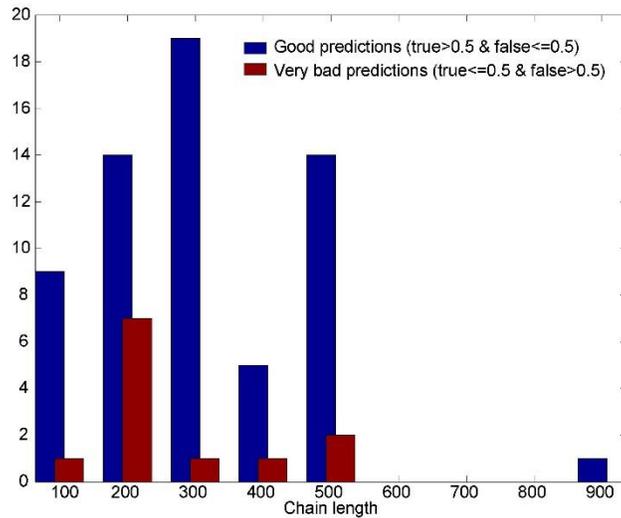

Figure A11. The prediction histogram over the sequence lengths for the prediction based on the adjustable number of fast modes and combined 1D & fixed 3D influence per hot residue for chains in dimers with high sequence length ratio (length ratio > 2, length > 80 residues). The influence is first spread linearly, upstream and downstream along the sequence, and then the it is spread to residue's spatial neighbors, the ones closer than 6 or 8 Å). True positives mean is 56.85 %, and the false positives mean is 43.49 %. There is 60.78 % of good predictions and 11.76 % of very bad predictions.

Full protein list (all 414 dimers) – to be added.